\documentclass[12pt]{article}
\usepackage{cite,amsmath,epsfig}
\usepackage{graphicx}
\usepackage{amssymb}

\newcommand{\nn}{\nonumber}
\newcommand{\eq}[1]{eq.~(\ref{#1})}
\newcommand{\Eq}[1]{eq.~(\ref{#1})}
\usepackage{cite}
\newcommand{\bea}{\begin{eqnarray}}
\newcommand{\eea}{\end{eqnarray}}

\newcommand{\complex}{{{\rm I} \kern -.59em {\rm C}}}

\def\citere#1{\mbox{Ref.~\cite{#1}}}

\def\beq{\begin{equation}}
\def\eeq{\end{equation}}

\newcommand{\FUTA}{{\bf FUTA}}
\newcommand{\FUTB}{{\bf FUTB}}
\def\refeq#1{\mbox{eq.~(\ref{#1})}}

\def\reffi#1{\mbox{Fig.~\ref{#1}}}

\def\citere#1{\mbox{Ref.~\cite{#1}}}

\newcommand{\MZ}{M_Z}

\newcommand{\tsf}{\theta\kern-.20em_{\tilde{f}}}
\newcommand{\tsfp}{\theta\kern-.20em_{\tilde{f}\prime}}
\newcommand{\tsq}{\theta\kern-.15em_{\tilde{q}}}

\newcommand{\VL}{\left( \begin{array}{c}}
\newcommand{\VR}{\end{array} \right)}
\newcommand{\ML}{\left( \begin{array}{cc}}
\newcommand{\MLd}{\left( \begin{array}{ccc}}
\newcommand{\MLv}{\left( \begin{array}{cccc}}
\newcommand{\MR}{\end{array} \right)}

\newcommand{\gev}{\,\, \mathrm{GeV}}

\newcommand{\BC}{\begin{center}}
\newcommand{\EC}{\end{center}}
\newcommand{\BE}{\begin{equation}}
\newcommand{\EE}{\end{equation}}
\newcommand{\BEA}{\begin{eqnarray}}
\newcommand{\BEAnn}{\begin{eqnarray*}}
\newcommand{\EEA}{\end{eqnarray}}
\newcommand{\EEAnn}{\end{eqnarray*}}

\newcommand{\id}{{\rm 1\kern-.12em
\rule{0.3pt}{1.5ex}\raisebox{0.0ex}{\rule{0.1em}{0.3pt}}}}

\def\si{\sigma}

\newcommand{\br}{{\rm BR}}

\newcommand{\amu}{a_\mu}

\newcommand{\amuexp}{a_\mu^{\rm exp}}
\newcommand{\amutheo}{a_\mu^{\rm theo}}

\begin{document}

\title{Finite Unification: Theory, Models and Predictions}

\author{
S. Heinemeyer\\
\small\em Instituto de F\'{\i}sica de Cantabria (CSIC-UC)\\[-1mm]
\small\em E-39005 Santander, Spain
\\[3mm]
M. Mondrag\'on \\
\small\em Instituto de F\'{\i}sica\\[-1mm]
\small\em Universidad Nacional Aut\'onoma de M\'exico\\[-1mm]
\small\em Apdo. Postal 20-364, M\'exico 01000
\\[3mm]
and \\[3mm]
 G. Zoupanos\footnote{On leave from Physics Deptartment, Nat. Technical University, 157 80 Zografou, Athens, Greece} \\
\small\em Institut f\"ur Theoretische Physik\\[-1mm]
\small\em Universit\"at Heidelberg\\[-1mm]
\small\em Philosophenweg 16\\[-1mm]
\small\em D-69120 Heidelberg, Germany}
\date{}
\maketitle

\begin{abstract}
All-loop Finite Unified Theories (FUTs) are very interesting $N=1$ 
supersymmetric Grand Unified Theories (GUTs)  realising an 
old field theory dream, and moreover have a remarkable predictive power due 
to the required reduction of couplings. The reduction of the dimensionless 
couplings in $N=1$ GUTs is achieved by searching for renormalization group 
invariant (RGI) relations among them holding beyond the unification 
scale. Finiteness results from the fact that there exist RGI relations 
among dimensional couplings that guarantee the vanishing of all 
beta-functions in certain $N=1$ GUTs even to all orders. Furthermore 
developments in the soft supersymmetry breaking sector of $N=1$ GUTs and
FUTs lead to exact RGI relations, i.e. reduction of couplings, in this 
dimensionful sector of the theory, too. Based on the above theoretical 
framework phenomenologically consistent FUTs have been constructed. Here 
we review FUT models based on the $SU(5)$ and $SU(3)^3$ gauge groups and 
their predictions. Of particular interest is the Higgs mass prediction of 
the models which is expected to be tested at the LHC.
\end{abstract}


\section{Introduction}

A large and sustained effort has been done in the recent years aiming
to achieve a unified description of all interactions. Out of this
endeavor two main directions have emerged as the most promising to
attack the problem, namely, the superstring theories and 
non-commutative geometry. The two approaches, although at a different
stage of development, have common unification targets and share similar
hopes for exhibiting improved renormalization properties in the
ultraviolet (UV) as compared to ordinary field theories.  Moreover the
two frameworks came closer by the observation that a natural
realization of non-commutativity of space appears in the string theory
context of D-branes in the presence of a constant background
antisymmetric field \cite{Connes:1997cr}. 
Among the numerous important developments in both frameworks, it is
worth noting two conjectures of utmost importance that signal the
developments in certain directions in string theory and not only,
related to the main theme of the present
review. The conjectures refer to 
(i) the duality among the 4-dimensional $N=4$ supersymmetric
Yang-Mills theory and the type IIB string theory on $AdS_5 \times S^5$
\cite{Maldacena:1997re}; the former being the maximal $N=4$
supersymmetric Yang-Mills theory is known to
be UV all-loop finite theory \cite{Mandelstam:1982cb,Brink:1982wv}, 
(ii) the possibility of ``miraculous'' UV divergence cancellations in
4-dimensional maximal $N=8$ supergravity leading to a finite theory,
as has been recently confirmed in a remarkable 4-loop calculation
\cite{Bern:2009kd,Kallosh:2009jb,Bern:2007hh,Bern:2006kd,Green:2006yu}.
However, despite the importance of having frameworks to discuss
quantum gravity in a self-consistent way and possibly to construct
there finite theories, it is also very interesting to search for the
minimal realistic framework in which finiteness can take place. 
After all the history of our field teaches us that if a new idea 
works, it does that in its simplest form.
 In
addition, the main goal expected from a unified description of
interactions by the particle physics community is to understand the
present day large number of free parameters of the Standard Model (SM)
in terms of a few fundamental ones. In other words, to achieve {\it
  reduction of couplings} at a more fundamental level. 

To reduce the number of free parameters of a theory, and thus render
it more predictive, one is usually led to introduce a symmetry.  Grand
Unified Theories (GUTs) are very good examples of such a procedure
\cite
{Pati:1973rp,Georgi:1974sy,Georgi:1974yf,Fritzsch:1974nn,Carlson:1975gu}.
For instance, in the case of minimal $SU(5)$, because of (approximate)
gauge coupling unification, it was possible to reduce the gauge
couplings by one and give a prediction for one of them. 
In fact, LEP data \cite{Amaldi:1991cn} 
seem to
suggest that a further symmetry, namely $N=1$ global supersymmetry (SUSY) 
\cite{Dimopoulos:1981zb,Sakai:1981gr} 
should also be required to make the prediction viable.
GUTs can also
relate the Yukawa couplings among themselves, again $SU(5)$ provided
an example of this by predicting the ratio $M_{\tau}/M_b$
\cite{Buras:1977yy} in the SM.  Unfortunately, requiring
more gauge symmetry does not seem to help, since additional
complications are introduced due to new degrees of freedom, in the
ways and channels of breaking the symmetry, and so on.

A natural extension of the GUT idea is to find a way to relate the
gauge and Yukawa sectors of a theory, that is to achieve Gauge-Yukawa
Unification (GYU) \cite{Kubo:1995cg,Kubo:1997fi,Kobayashi:1999pn}.  A
symmetry which naturally relates the two sectors is supersymmetry, in
particular $N=2$ SUSY \cite{Fayet:1978ig}.  It turns out, however, that $N=2$
supersymmetric theories have serious phenomenological problems due to
light mirror fermions.  Also in superstring theories and in composite
models there exist relations among the gauge and Yukawa couplings, but
both kind of theories have phenomenological problems, which we are not
going to address here.


In our studies
\cite{Kubo:1995cg,Kubo:1997fi,Kobayashi:1999pn,Kapetanakis:1992vx,Mondragon:1993tw,Kubo:1994bj,Kubo:1994xa,Kubo:1995zg,Kubo:1996js}
we have developed a complementary strategy in searching for a more
fundamental theory possibly at the Planck scale, whose basic
ingredients are GUTs and supersymmetry, but its consequences certainly
go beyond the known ones. Our method consists of hunting for
renormalization group invariant (RGI) relations holding below the
Planck scale, which in turn are preserved down to the GUT scale. This
programme, called Gauge--Yukawa unification scheme, applied in the
dimensionless couplings of supersymmetric GUTs, such as gauge and
Yukawa couplings, had already noticable successes by predicting
correctly, among others, the top quark mass in the finite and in the
minimal $N = 1$ supersymmetric SU(5) GUTs
\cite{Kapetanakis:1992vx,Mondragon:1993tw,Kubo:1994bj}.
An impressive aspect of the RGI relations is that one can guarantee
their validity to all-orders in perturbation theory by studying the
uniqueness of the resulting relations at one-loop, as was proven in
the early days of the programme of {\it reduction of couplings}
\cite{Zimmermann:1984sx,Oehme:1984yy,Ma:1977hf,Ma:1984by}. Even more
remarkable is the fact that it is possible to find RGI relations among
couplings that guarantee finiteness to all-orders in perturbation
theory
\cite{Lucchesi:1987ef,Lucchesi:1987he,Lucchesi:1996ir,Ermushev:1986cu,Kazakov:1987vg}.

It is worth noting that the above principles have only  been applied in
supersymmetric GUTs for reasons that will be transparent in the
following sections.  We should also stress that our conjecture for GYU
is by no means in conflict with the interesting proposals mentioned
before (see also
ref.\cite{Schrempp:1992zi,Schrempp:1994xn,Schrempp:1996fb}), but it
rather uses all of them, hopefully in a more successful perspective.
For instance, the use of SUSY GUTs comprises the demand of the
cancellation of quadratic divergences in the SM.  Similarly, the very
interesting conjectures about the infrared fixed points are
generalized in our proposal, since searching for RGI relations among
various couplings corresponds to searching for {\it fixed points} of
the coupled differential equations obeyed by the various couplings of
a theory.

Although SUSY seems to be an essential feature for a
successful realization of the above programme, its breaking has to be
understood too, since it has the ambition to supply the SM with
predictions for several of its free parameters. Indeed, the search for
RGI relations has been extended to the soft SUSY breaking
sector (SSB) of these theories \cite{Kubo:1996js,Jack:1995gm}, which
involves parameters of dimension one and two.  Then a very interesting
progress has been made
\cite{Hisano:1997ua,Jack:1997pa,Avdeev:1997vx,Kazakov:1998uj,Kazakov:1997nf,Jack:1997eh,Kobayashi:1998jq}
concerning the renormalization properties of the SSB parameters based
conceptually and technically on the work of
ref.~\cite{Yamada:1994id}: the powerful
supergraph method
\cite{Delbourgo:1974jg,Salam:1974pp,Fujikawa:1974ay,Grisaru:1979wc}
for studying supersymmetric theories has been applied to the softly
broken ones by using the ``spurion'' external space-time independent
superfields \cite{Girardello:1981wz}.  In the latter method a softly
broken supersymmetric gauge theory is considered as a supersymmetric
one in which the various parameters such as couplings and masses have
been promoted to external superfields that acquire ``vacuum
expectation values''. Based on this method the relations among the
soft term renormalization and that of an unbroken supersymmetric
theory have been derived. In particular the $\beta$-functions of the
parameters of the softly broken theory are expressed in terms of
partial differential operators involving the dimensionless parameters
of the unbroken theory. The key point in the strategy of
refs.~\cite{Kazakov:1998uj,Kazakov:1997nf,Jack:1997eh,Kobayashi:1998jq}
in solving the set of coupled differential equations so as to be able
to express all parameters in a RGI way, was to transform the partial
differential operators involved to total derivative operators.  This
is indeed possible to be done on the RGI surface which is defined by
the solution of the reduction equations.

  On the phenomenological side there exist some serious developments, 
  too.  Previously an appealing ``universal'' set of soft scalar
  masses was asummed in the SSB sector of supersymmetric theories,
  given that apart from economy and simplicity (1) they are part of
  the constraints that preserve finiteness up to two-loops
  \cite{Jones:1984cu,Jack:1994kd}, (2) they are RGI up to two-loops in
  more general supersymmetric gauge theories, subject to the condition
  known as $P =1/3~Q$ \cite{Jack:1995gm} and (3) they appear in the
  attractive dilaton dominated SUSY breaking superstring
  scenarios \cite{Ibanez:1992hc,Kaplunovsky:1993rd,Brignole:1993dj}.
  However, further studies have exhibited a number of problems all due
  to the restrictive nature of the ``universality'' assumption for the
  soft scalar masses.  For instance, (a) in finite unified theories the
  universality predicts that the lightest supersymmetric particle is a
  charged particle, namely the superpartner of the $\tau$ lepton
  $\tilde\tau$, (b) the MSSM with universal soft scalar masses is
  inconsistent with the attractive radiative electroweak symmetry
  breaking \cite{Brignole:1993dj}, and (c) which is the worst of all,
  the universal soft scalar masses lead to charge and/or colour
  breaking minima deeper than the standard vacuum \cite{Casas:1996wj}. 
  Therefore, there have been attempts to relax this constraint without
  loosing its attractive features. First an interesting observation
  was made that in $N = 1$ Gauge--Yukawa unified theories there exists
  a RGI sum rule for the soft scalar masses at lower orders; at
  one-loop for the non-finite case \cite{Kawamura:1997cw} and at
  two-loops for the finite case \cite{Kobayashi:1997qx}. The sum rule
  manages to overcome the above unpleasant phenomenological
  consequences. Moreover it was proven \cite{Kobayashi:1998jq} that
  the sum rule for the soft scalar massses is RGI to all-orders for
  both the general as well as for the finite case. Finally, the exact
  $\beta$-function for the soft scalar masses in the
  Novikov-Shifman-Vainstein-Zakharov (NSVZ) scheme
  \cite{Novikov:1983ee,Novikov:1985rd,Shifman:1996iy} for the softly
  broken supersymmetric QCD has been obtained \cite{Kobayashi:1998jq}.
  Armed with the above tools and results we are in a position to study
  the spectrum of the full finite models in terms of few free
  parameters with emphasis on \break
 the predictions for the lightest Higgs
  mass, which is expected to be tested at LHC.

\section{Unification of Couplings by the RGI Method}

Let us next briefly outline the idea of reduction of couplings.  
Any RGI relation among couplings 
(which does not depend on the renormalization
scale $\mu$ explicitly) can be expressed,
in the implicit form $\Phi (g_1,\cdots,g_A) ~=~\mbox{const.}$,
which
has to satisfy the partial differential equation (PDE)
\bea
\mu\,\frac{d \Phi}{d \mu} &=& {\vec \nabla}\cdot {\vec \beta} ~=~ 
\sum_{a=1}^{A} 
\,\beta_{a}\,\frac{\partial \Phi}{\partial g_{a}}~=~0~,
\eea
where $\beta_a$ is the $\beta$-function of $g_a$.
This PDE is equivalent
to a set of ordinary differential equations, 
the so-called reduction equations (REs) \cite{Zimmermann:1984sx,Oehme:1984yy,Oehme:1985jy},
\bea
\beta_{g} \,\frac{d g_{a}}{d g} &=&\beta_{a}~,~a=1,\cdots,A~,
\label{redeq}
\eea
where $g$ and $\beta_{g}$ are the primary 
coupling and its $\beta$-function,
and the counting on $a$ does not include $g$.
Since maximally ($A-1$) independent 
RGI ``constraints'' 
in the $A$-dimensional space of couplings
can be imposed by the $\Phi_a$'s, one could in principle
express all the couplings in terms of 
a single coupling $g$.
 The strongest requirement is to demand
 power series solutions to the REs,
\bea
g_{a} &=& \sum_{n}\rho_{a}^{(n)}\,g^{2n+1}~,
\label{powerser}
\eea which formally preserve perturbative renormalizability.
Remarkably, the uniqueness of such power series solutions can be
decided already at the one-loop level
\cite{Zimmermann:1984sx,Oehme:1984yy,Oehme:1985jy}.  To illustrate
this, let us assume that the $\beta$-functions have the form \bea
\beta_{a} &=&\frac{1}{16 \pi^2}[ \sum_{b,c,d\neq
  g}\beta^{(1)\,bcd}_{a}g_b g_c g_d+
\sum_{b\neq g}\beta^{(1)\,b}_{a}g_b g^2]+\cdots~,\nn\\
\beta_{g} &=&\frac{1}{16 \pi^2}\beta^{(1)}_{g}g^3+ \cdots~, \eea where
$\cdots$ stands for higher order terms, and $ \beta^{(1)\,bcd}_{a}$'s
are symmetric in $ b,c,d$.  We then assume that the $\rho_{a}^{(n)}$'s
with $n\leq r$ have been uniquely determined. To obtain
$\rho_{a}^{(r+1)}$'s, we insert the power series (\ref{powerser}) into
the REs (\ref{redeq}) and collect terms of ${\cal O}(g^{2r+3})$ and
find 
\bea 
\sum_{d\neq g}M(r)_{a}^{d}\,\rho_{d}^{(r+1)} &=& \mbox{lower
  order quantities}~,\nn 
\eea 
where the r.h.s. is known by assumption,
and 
\bea 
M(r)_{a}^{d} &=&3\sum_{b,c\neq
  g}\,\beta^{(1)\,bcd}_{a}\,\rho_{b}^{(1)}\,
\rho_{c}^{(1)}+\beta^{(1)\,d}_{a}
-(2r+1)\,\beta^{(1)}_{g}\,\delta_{a}^{d}~,\label{M}\\
0 &=&\sum_{b,c,d\neq g}\,\beta^{(1)\,bcd}_{a}\,
\rho_{b}^{(1)}\,\rho_{c}^{(1)}\,\rho_{d}^{(1)} +\sum_{d\neq
  g}\beta^{(1)\,d}_{a}\,\rho_{d}^{(1)}
-\beta^{(1)}_{g}\,\rho_{a}^{(1)}~, 
\eea 

 Therefore, the $\rho_{a}^{(n)}$'s for all $n > 1$ for a
given set of $\rho_{a}^{(1)}$'s can be uniquely determined if $\det
M(n)_{a}^{d} \neq 0$ for all $n \geq 0$.

As it will be clear later by examining specific examples, the various
couplings in supersymmetric theories have easily the same asymptotic
behaviour.  Therefore searching for a power series solution of the
form (\ref{powerser}) to the REs (\ref{redeq}) is justified. This is
not the case in non-supersymmetric theories, although the deeper
reason for this fact is not fully understood.

The possibility of coupling unification described in this section  
is without any doubt
attractive because the ``completely reduced'' theory contains 
only one independent coupling, but  it can be
unrealistic. Therefore, one often would like to impose fewer RGI
constraints, and this is the idea of partial reduction \cite{Kubo:1985up,Kubo:1988zu}.

\section{Reduction of dimensionful parameters}

The reduction of couplings
 was originally formulated for massless theories
on the basis of the Callan-Symanzik equation \cite{Zimmermann:1984sx,Oehme:1984yy,Oehme:1985jy}.
The extension to theories with massive parameters
is not straightforward if one wants to keep
the generality and the rigor
on the same level as for the massless case;
one has 
to fulfill a set of requirements coming from
the renormalization group
equations,  the  Callan-Symanzik equations, etc.
along with the normalization
conditions imposed on irreducible Green's functions \cite{Piguet:1989pc}.
See \cite{Zimmermann:2000hn} for  interesting results in this direction. 
Here, to simplify the situation,  we would like  to
 assume  that
 a mass-independent renormalization scheme has been
employed so that all the  RG functions have only  trivial
dependencies of dimensional parameters. 

To be general, we consider  a renormalizable theory
which contain a set of $(N+1)$ dimension-zero couplings,
$\{\hat{g}_0,\hat{g}_1,\dots,\hat{g}_N\}$, a set of $L$ 
parameters with  dimension 
one, $\{\hat{h}_1,\dots,\hat{h}_L\}$,
and a set of $M$ parameters with dimension two,
$\{\hat{m}_{1}^{2},\dots,\hat{m}_{M}^{2}\}$. 
The renormalized irreducible vertex function 
satisfies the RG equation
\bea
0 &=& {\cal D}\Gamma [~{\bf
\Phi}'s;\hat{g}_0,\hat{g}_1,\dots,\hat{g}_N;\hat{h}_1,\dots,\hat{h}_L;
\hat{m}^{2}_{1},\dots,\hat{m}^{2}_{M};\mu~]~, \label{vertex}\\
{\cal D} &=& \mu\frac{\partial}{\partial \mu}+
~\sum_{i=0}^{N}\,\beta_i\,
\frac{\partial}{\partial \hat{g}_i}+
\sum_{a=1}^{L}\,\gamma_{a}^{h}\,
\frac{\partial}{\partial \hat{h}_a}+
\sum_{\alpha=1}^{M}\,\gamma^{m^2}_{\alpha}\frac{\partial}
{\partial \hat{m}_{\alpha}^{2}}+ ~\sum_{J}\,\Phi_I
\gamma^{\phi I}_{~~~J} \frac{\delta}{\delta \Phi_J}~.\nn
\eea
Since we assume a mass-independent renormalization scheme,
the $\gamma$'s have the form
\bea
\gamma_{a}^{h} &=& \sum_{b=1}^{L}\,
\gamma_{a}^{h,b}(g_0,\dots,g_N)\hat{h}_b~,\nn\\
\gamma_{\alpha}^{m^2} &=&
\sum_{\beta=1}^{M}\,\gamma_{\alpha}^{m^2,\beta}(g_0,\dots,g_N)
\hat{m}_{\beta}^{2}+
\sum_{a,b=1}^{L}\,\gamma_{\alpha}^{m^2,a b}
(g_0,\dots,g_N)\hat{h}_a \hat{h}_b~,
\eea
where $\gamma_{a}^{h,b}, 
\gamma_{\alpha}^{m^2,\beta}$ and 
  $\gamma_{a}^{m^2,a b}$
are power series of the dimension-zero
couplings $g$'s in perturbation theory.

As in the massless case, we then look for 
 conditions under which the reduction of
parameters,
\bea
\hat{g}_i &=&\hat{g}_i(g)~,~(i=1,\dots,N)~,\label{gr}\\
~\hat{h}_a &= &\sum_{b=1}^{P}\,
f_{a}^{b}(g) h_b~,~(a=P+1,\dots,L)~,\label{h}\\
~\hat{m}_{\alpha}^{2} &= &\sum_{\beta=1}^{Q}\,
e_{\alpha}^{\beta}(g) m_{\beta}^{2}+
\sum_{a,b=1}^{P}\,k_{\alpha}^{a b}(g)
h_a h_b~,~(\alpha=Q+1,\dots,M)~,\label{m}
\eea
is consistent with the RG equation (1),
where we assume that $g\equiv g_0$, $h_a \equiv
\hat{h}_a~~(1 \leq a \leq P)$ and 
$m_{\alpha}^{2} \equiv
\hat{m}_{\alpha}^{2}~~(1 \leq \alpha \leq Q)$ 
are independent parameters of the reduced theory.
We find  that the following set of
equations has to be satisfied:
\bea
\beta_g\,\frac{\partial
\hat{g}_{i}}{\partial g} & =& \beta_i ~,~(i=1,\dots,N)~, \label{betagr}\\
\beta_g\,\frac{\partial
\hat{h}_{a}}{\partial g}+\sum_{b=1}^{P}\gamma_{b}^{h}
\frac{\partial
\hat{h}_{a}}{\partial
h_b} &=&\gamma_{a}^{h}~,~(a=P+1,\dots,L)~,\label{betah}\\
\beta_g\,\frac{\partial
\hat{m}^{2}_{\alpha}}{\partial g}
+\sum_{a=1}^{P}\gamma_{a}^{h}
\frac{\partial
\hat{m}^{2}_{\alpha}}{\partial
h_a}+
\sum_{\beta=1}^{Q}\gamma_{\beta}^{m^2}
\frac{\partial
\hat{m}^{2}_{\alpha}}{\partial
m^{2}_{\beta}}
 &=&\gamma_{\alpha}^{m^2}~,~(\alpha=Q+1,\dots,M)~. \label{betam}
\eea
Using eq.(\ref{vertex}) for $\gamma$'s, one finds that
eqs.(\ref{betagr}-\ref{betam}) 
reduce to 
\bea
& &\beta_g\,\frac{d f_{a}^{b}}{d g}+
\sum_{c=1}^{P}\, f_{a}^{c} 
[\,\gamma_{c}^{h,b}+\sum_{d=P+1}^{L}\,
\gamma_{c}^{h,d}f_{d}^{ b}\,]
-\gamma_{a}^{h,b}-\sum_{d=P+1}^{L}\,
\gamma_{a}^{h,d}f_{d}^{ b}~=0~,\label{red1}\\
& &~(a=P+1,\dots,L; b=1,\dots,P)~,\nn\\
& &\beta_g\,\frac{d e_{\alpha}^{\beta}}{d g}+
\sum_{\gamma=1}^{Q}\, e_{\alpha}^{\gamma} 
[\,\gamma_{\gamma}^{m^2,\beta}+\sum_{\delta=Q+1}^{M}\,
\gamma_{\gamma}^{m^2,\delta}e_{\delta}^{\beta}\,]
-\gamma_{\alpha}^{m^2,\beta}-\sum_{\delta=Q+1}^{M}\,
\gamma_{\alpha}^{m^2,\delta}e_{\delta}^{\beta}~=0~,\label{red2}\\
& &~(\alpha=Q+1,\dots,M; \beta=1,\dots,Q)~,\nn\\
& &\beta_g\,\frac{d k_{\alpha}^{a b}}{d g}+2\sum_{c=1}^{P}\,
(\,\gamma_{c}^{h,a}+\sum_{d=P+1}^{L}\,
\gamma_{c}^{h,d}f_{d}^{a}\,)k_{\alpha}^{c b}+
\sum_{\beta=1}^{Q}\, e_{\alpha}^{\beta}
[\,\gamma_{\beta}^{m^2,a b}+\sum_{c,d=P+1}^{L}\,
\gamma_{\beta}^{m^2,c d}f_{c}^{a} f_{d}^{b}\nn\\
& &+2\sum_{c=P+1}^{L}\,\gamma_{\beta}^{m^2,c b}f_{c}^{a}+
\sum_{\delta=Q+1}^{M}\,\gamma_{\beta}^{m^2,\delta}
k_{\delta}^{a b} \,]-
[\,\gamma_{\alpha}^{m^2,a b}+\sum_{c,d=P+1}^{L}\,
\gamma_{\alpha}^{m^2,c d}f_{c}^{a} f_{d}^{b}\nn\\
& &+
2\sum_{c=P+1}^{L}\,\gamma_{\alpha}^{m^2,c b}f_{c}^{a}
+\sum_{\delta=Q+1}^{M}\,\gamma_{\alpha}^{m^2,\delta}
k_{\delta}^{a b} \,]~=0~,\label{red3}\\
& &(\alpha=Q+1,\dots,M; a,b=1,\dots,P)~.\nn
\eea
If these equations are satisfied, 
the irreducible vertex function of the reduced theory
\bea
& &\Gamma_R [~{\bf
\Phi}'s; g; h_1,\dots,h_P; m^{2}_{1},
\dots,\hat{m}^{2}_{Q};\mu~]~\nn\\
&\equiv& \Gamma [~{\bf
\Phi}'s; g,\hat{g}_1(g),\dots,\hat{g}_N (g);
 h_1,\dots,h_P, \hat{h}_{P+1}(g,h),\dots,\hat{h}_L(g,h);\nn\\
& & m^{2}_{1},\dots,\hat{m}^{2}_{Q},\hat{m}^{2}_{Q+1}(g,h,m^2),
\dots,\hat{m}^{2}_{M}(g,h,m^2);\mu~] 
\eea
has
the same renormalization group flow as the original one.

The requirement for the reduced theory to be perturbative
renormalizable means that the functions $\hat{g}_i $, $f_{a}^{b} $,
$e_{\alpha}^{\beta}$ and $k_{\alpha}^{a b}$, defined in
eqs.~(\ref{gr}-\ref{m}), should have a power series expansion in the
primary coupling $g$: 
\bea 
\hat{g}_{i} &=& g\,\sum_{n=0}^{\infty}
\rho_{i}^{(n)} g^{n}~,~
f_{a}^{b}= g\sum_{n=0}^{\infty} \eta_{a}^{b~(n)} g^{n}~,\nn\\~
e_{\alpha}^{\beta} &= &\sum_{n=0}^{\infty} \xi_{\alpha}^{\beta~(n)}
g^{n}~,~ k_{\alpha}^{a b }= \sum_{n=0}^{\infty} \chi_{\alpha}^{a
  b~(n)} g^{n}~. 
\eea 
To obtain the expansion coefficients, we insert
the power series ansatz above into
eqs.~(\ref{betagr},\ref{red1}--\ref{red3}) and require that the
equations are satisfied at each order in $g$. Note that the existence
of a unique power series solution is a non-trivial matter: It depends
on the theory as well as on the choice of the set of independent
parameters.

\section{ Finiteness in N=1 Supersymmetric Gauge Theories}
\label{sec:futs}

Let us consider a chiral, anomaly free,
$N=1$ globally supersymmetric
gauge theory based on a group G with gauge coupling
constant $g$. The
superpotential of the theory is given by
\bea
W&=& \frac{1}{2}\,m_{ij} \,\phi_{i}\,\phi_{j}+
\frac{1}{6}\,C_{ijk} \,\phi_{i}\,\phi_{j}\,\phi_{k}~,
\label{supot}
\eea
where $m_{ij}$ and $C_{ijk}$ are gauge invariant tensors and
the matter field $\phi_{i}$ transforms
according to the irreducible representation  $R_{i}$
of the gauge group $G$. The
renormalization constants associated with the
superpotential (\ref{supot}), assuming that
SUSY is preserved, are
\bea
\phi_{i}^{0}&=&(Z^{j}_{i})^{(1/2)}\,\phi_{j}~,~\\
m_{ij}^{0}&=&Z^{i'j'}_{ij}\,m_{i'j'}~,~\\
C_{ijk}^{0}&=&Z^{i'j'k'}_{ijk}\,C_{i'j'k'}~.
\eea
The $N=1$ non-renormalization theorem \cite{Wess:1973kz,Iliopoulos:1974zv,Fujikawa:1974ay} ensures that
there are no mass
and cubic-interaction-term infinities and therefore
\bea
Z_{ijk}^{i'j'k'}\,Z^{1/2\,i''}_{i'}\,Z^{1/2\,j''}_{j'}
\,Z^{1/2\,k''}_{k'}&=&\delta_{(i}^{i''}
\,\delta_{j}^{j''}\delta_{k)}^{k''}~,\nn\\
Z_{ij}^{i'j'}\,Z^{1/2\,i''}_{i'}\,Z^{1/2\,j''}_{j'}
&=&\delta_{(i}^{i''}
\,\delta_{j)}^{j''}~.
\eea
As a result the only surviving possible infinities are
the wave-function renormalization constants
$Z^{j}_{i}$, i.e.,  one infinity
for each field. The one -loop $\beta$-function of the gauge
coupling $g$ is given by \cite{Parkes:1984dh}
\bea
\beta^{(1)}_{g}=\frac{d g}{d t} =
\frac{g^3}{16\pi^2}\,[\,\sum_{i}\,l(R_{i})-3\,C_{2}(G)\,]~,
\label{betag}
\eea
where $l(R_{i})$ is the Dynkin index of $R_{i}$ and $C_{2}(G)$
 is the
quadratic Casimir of the adjoint representation of the
gauge group $G$. The $\beta$-functions of
$C_{ijk}$,
by virtue of the non-renormalization theorem, are related to the
anomalous dimension matrix $\gamma_{ij}$ of the matter fields
$\phi_{i}$ as:
\beq
\beta_{ijk} =
 \frac{d C_{ijk}}{d t}~=~C_{ijl}\,\gamma^{l}_{k}+
 C_{ikl}\,\gamma^{l}_{j}+
 C_{jkl}\,\gamma^{l}_{i}~.
\label{betay}
\eeq
At one-loop level $\gamma_{ij}$ is \cite{Parkes:1984dh}
\beq
\gamma^{i(1)}_j=\frac{1}{32\pi^2}\,[\,
C^{ikl}\,C_{jkl}-2\,g^2\,C_{2}(R_{i})\delta_{j}^1\,],
\label{gamay}
\eeq
where $C_{2}(R_{i})$ is the quadratic Casimir of the representation
$R_{i}$, and $C^{ijk}=C_{ijk}^{*}$.
Since
dimensional coupling parameters such as masses  and couplings of cubic
scalar field terms do not influence the asymptotic properties 
 of a theory on which we are interested here, it is
sufficient to take into account only the dimensionless supersymmetric
couplings such as $g$ and $C_{ijk}$.
So we neglect the existence of dimensional parameters, and
assume furthermore that
$C_{ijk}$ are real so that $C_{ijk}^2$ always are positive numbers.

As one can see from Eqs.~(\ref{betag}) and (\ref{gamay}),
 all the one-loop $\beta$-functions of the theory vanish if
 $\beta_g^{(1)}$ and $\gamma _{ij}^{(1)}$ vanish, i.e.
\begin{equation}
\sum _i \ell (R_i) = 3 C_2(G) \,,
\label{1st}
\end{equation}

\begin{equation}
C^{ikl} C_{jkl} = 2\delta ^i_j g^2  C_2(R_i)\,,
\label{2nd}
\end{equation}

The conditions for finiteness for $N=1$ field theories with $SU(N)$ gauge
symmetry are discussed in \cite{Rajpoot:1984zq}, and the
analysis of the anomaly-free and no-charge renormalization
requirements for these theories can be found in \cite{Rajpoot:1985aq}. 
A very interesting result is that the conditions (\ref{1st},\ref{2nd})
are necessary and sufficient for finiteness at the two-loop level
\cite{Parkes:1984dh,West:1984dg,Jones:1985ay,Jones:1984cx,Parkes:1985hh}.

In case SUSY is broken by soft terms, the requirement of
finiteness in the one-loop soft breaking terms imposes further
constraints among themselves \cite{Jones:1984cu}.  In addition, the same set
of conditions that are sufficient for one-loop finiteness of the soft
breaking terms render the soft sector of the theory two-loop finite
\cite{Jones:1984cu}.

The one- and two-loop finiteness conditions (\ref{1st},\ref{2nd}) restrict
considerably the possible choices of the irreps. $R_i$ for a given
group $G$ as well as the Yukawa couplings in the superpotential
(\ref{supot}).  Note in particular that the finiteness conditions cannot be
applied to the minimal supersymmetric standard model (MSSM), since the presence
of a $U(1)$ gauge group is incompatible with the condition
(\ref{1st}), due to $C_2[U(1)]=0$.  This naturally leads to the
expectation that finiteness should be attained at the grand unified
level only, the MSSM being just the corresponding, low-energy,
effective theory.

Another important consequence of one- and two-loop finiteness is that
SUSY (most probably) can only be broken due to the soft
breaking terms.  Indeed, due to the unacceptability of gauge singlets,
F-type spontaneous symmetry breaking \cite{O'Raifeartaigh:1975pr}
terms are incompatible with finiteness, as well as D-type
\cite{Fayet:1974jb} spontaneous breaking which requires the existence
of a $U(1)$ gauge group.

A natural question to ask is what happens at higher loop orders.  The
answer is contained in a theorem
\cite{Lucchesi:1987he,Lucchesi:1987ef} which states the necessary and
sufficient conditions to achieve finiteness at all orders.  Before we
discuss the theorem let us make some introductory remarks.  The
finiteness conditions impose relations between gauge and Yukawa
couplings.  To require such relations which render the couplings
mutually dependent at a given renormalization point is trivial.  What
is not trivial is to guarantee that relations leading to a reduction
of the couplings hold at any renormalization point.  As we have seen,
the necessary and also sufficient, condition for this to happen is to
require that such relations are solutions to the REs \beq \beta _g
\frac{d C_{ijk}}{dg} = \beta _{ijk}
\label{redeq2}
\eeq
and hold at all orders.   Remarkably, the existence of 
all-order power series solutions to (\ref{redeq2}) can be decided at
one-loop level, as already mentioned.

Let us now turn to the all-order finiteness theorem
\cite{Lucchesi:1987he,Lucchesi:1987ef}, which states that if a $N=1$
supersymmetric gauge theory can become finite to all orders in the
sense of vanishing $\beta$-functions, that is of physical scale
invariance.  It is based on (a) the structure of the supercurrent in
$N=1$ supersymmetric gauge theory
\cite{Ferrara:1974pz,Piguet:1981mu,Piguet:1981mw}, and on (b) the
non-renormalization properties of $N=1$ chiral anomalies
\cite{Lucchesi:1987he,Lucchesi:1987ef,Piguet:1986td,Piguet:1986pk,Ensign:1987wy}.
Details on the proof can be found in
refs. \cite{Lucchesi:1987he,Lucchesi:1987ef} and further discussion in
refs.~\cite{Piguet:1986td,Piguet:1986pk,Ensign:1987wy,Lucchesi:1996ir,Piguet:1996mx}.
Here, following mostly ref.~\cite{Piguet:1996mx} we present a
comprehensible sketch of the proof.
 
Consider a $N=1$ supersymmetric gauge theory, with simple Lie group
$G$.  The content of this theory is given at the classical level by
the matter supermultiplets $S_i$, which contain a scalar field
$\phi_i$ and a Weyl spinor $\psi_{ia}$, and the vector supermultiplet
$V_a$, which contains a gauge vector field $A_{\mu}^a$ and a gaugino
Weyl spinor $\lambda^a_{\alpha}$. 

Let us first recall certain facts about the theory:

\noindent (1)  A massless $N=1$ supersymmetric theory is invariant 
under a $U(1)$ chiral transformation $R$ under which the various fields 
transform as follows
\bea
A'_{\mu}&=&A_{\mu},~~\lambda '_{\alpha}=\exp({-i\theta})\lambda_{\alpha}\nn\\ 
\phi '&=& \exp({-i\frac{2}{3}\theta})\phi,~~\psi_{\alpha}'= \exp({-i\frac{1}
    {3}\theta})\psi_{\alpha},~\cdots
\eea
The corresponding axial Noether current $J^{\mu}_R(x)$ is
\beq
J^{\mu}_R(x)=\bar{\lambda}\gamma^{\mu}\gamma^5\lambda + \cdots
\label{noethcurr}
\eeq
is conserved classically, while in the quantum case is violated by the
axial anomaly
\beq
\partial_{\mu} J^{\mu}_R =
r(\epsilon^{\mu\nu\sigma\rho}F_{\mu\nu}F_{\sigma\rho}+\cdots).
\label{anomaly}
\eeq

From its known topological origin in ordinary gauge theories
\cite{AlvarezGaume:1983cs,Bardeen:1984pm,Zumino:1983rz}, one would expect that the axial vector current
$J^{\mu}_R$ to satisfy the Adler-Bardeen theorem  and
receive corrections only at the one-loop level.  Indeed it has been
shown that the same non-renormalization theorem holds also in
supersymmetric theories \cite{Piguet:1986td,Piguet:1986pk,Ensign:1987wy}.  Therefore
\beq
r=\hbar \beta_g^{(1)}.
\label{r}
\eeq

\noindent (2)  The massless theory we consider is scale invariant at
the classical level and, in general, there is a scale anomaly due to
radiative corrections.  The scale anomaly appears in the trace of the
energy momentum tensor $T_{\mu\nu}$, which is traceless classically.
It has the form
\bea
T^{\mu}_{\mu} &~=~& \beta_g F^{\mu\nu}F_{\mu\nu} +\cdots
\label{Tmm}
\eea

\noindent (3)  Massless, $N=1$ supersymmetric gauge theories are
classically invariant under the supersymmetric extension of the
conformal group -- the superconformal group.  Examining the
superconformal algebra, it can be seen that the subset of
superconformal transformations consisting of translations,
SUSY transformations, and axial $R$ transformations is closed
under SUSY, i.e. these transformations form a representation
of SUSY.  It follows that the conserved currents
corresponding to these transformations make up a supermultiplet
represented by an axial vector superfield called supercurrent
$J$,
\beq
J \equiv \{ J'^{\mu}_R, ~Q^{\mu}_{\alpha}, ~T^{\mu}_{\nu} , ... \},
\label{J}
\eeq
where $J'^{\mu}_R$ is the current associated to R invariance,
$Q^{\mu}_{\alpha}$ is the one associated to SUSY invariance,
and $T^{\mu}_{\nu}$ the one associated to translational invariance
(energy-momentum tensor). 

The anomalies of the R current $J'^{\mu}_R$, the trace
anomalies of the 
SUSY current, and the energy-momentum tensor, form also
a second supermultiplet, called the supertrace anomaly
\bea
S &=& \{ Re~ S, ~Im~ S,~S_{\alpha}\} =\nn\\
&& \{T^{\mu}_{\mu},~\partial _{\mu} J'^{\mu}_R,~\sigma^{\mu}_{\alpha
  \dot{\beta}} \bar{Q}^{\dot\beta}_{\mu}~+~\cdots \}
\eea
where $T^{\mu}_{\mu}$ in Eq.(\ref{Tmm}) and
\bea
\partial _{\mu} J'^{\mu}_R &~=~&\beta_g\epsilon^{\mu\nu\sigma\rho}
F_{\mu\nu}F_{\sigma\rho}+\cdots\\ 
\sigma^{\mu}_{\alpha \dot{\beta}} \bar{Q}^{\dot\beta}_{\mu}&~=~&\beta_g
\lambda^{\beta}\sigma^{\mu\nu}_{\alpha\beta}F_{\mu\nu}+\cdots 
\eea

\noindent (4) It is very important to note that 
the Noether current defined in (\ref{noethcurr}) is not the same as the
current associated to R invariance that appears in the
supercurrent 
$J$ in (\ref{J}), but they coincide in the tree approximation. 
So starting from a unique classical Noether current
$J^{\mu}_{R(class)}$,  the Noether
current $J^{\mu}_R$ is defined as the quantum extension of
$J^{\mu}_{R(class)}$ which allows for the
validity of the non-renormalization theorem.  On the other hand
$J'^{\mu}_R$, is defined to belong to the supercurrent $J$,
together with the energy-momentum tensor.  The two requirements
cannot be fulfilled by a single current operator at the same time.

Although the Noether current $J^{\mu}_R$ which obeys (\ref{anomaly})
and the current $J'^{\mu}_R$ belonging to the supercurrent multiplet
$J$ are not the same, there is a relation
\cite{Lucchesi:1987he,Lucchesi:1987ef} between quantities associated
with them 
\beq 
r=\beta_g(1+x_g)+\beta_{ijk}x^{ijk}-\gamma_Ar^A
\label{rbeta}
\eeq
where $r$ was given in Eq.~(\ref{r}).  The $r^A$ are the
non-renormalized coefficients of 
the anomalies of the Noether currents associated to the chiral
invariances of the superpotential, and --like $r$-- are strictly
one-loop quantities. The $\gamma_A$'s are linear
combinations of the anomalous dimensions of the matter fields, and
$x_g$, and $x^{ijk}$ are radiative correction quantities.
The structure of equality (\ref{rbeta}) is independent of the
renormalization scheme.

One-loop finiteness, i.e. vanishing of the $\beta$-functions at one-loop,
implies that the Yukawa couplings $\lambda_{ijk}$ must be functions of
the gauge coupling $g$. To find a similar condition to all orders it
is necessary and sufficient for the Yukawa couplings to be a formal
power series in $g$, which is solution of the REs (\ref{redeq2}).  

We can now state the theorem for all-order vanishing
$\beta$-functions.
\bigskip

{\bf Theorem:}

Consider an $N=1$ supersymmetric Yang-Mills theory, with simple gauge
group. If the following conditions are satisfied
\begin{enumerate}
\item There is no gauge anomaly.
\item The gauge $\beta$-function vanishes at one-loop
  \beq
  \beta^{(1)}_g = 0 =\sum_i l(R_{i})-3\,C_{2}(G).
  \eeq
\item There exist solutions of the form
  \beq
  C_{ijk}=\rho_{ijk}g,~\qquad \rho_{ijk}\in\complex
  \label{soltheo}
  \eeq
to the  conditions of vanishing one-loop matter fields anomalous dimensions
  \bea
  &&\gamma^{i~(1)}_j~=~0\\
  &&=\frac{1}{32\pi^2}~[ ~
  C^{ikl}\,C_{jkl}-2~g^2~C_{2}(R_{i})\delta_{ij} ].\nn
  \eea
\item these solutions are isolated and non-degenerate when considered
  as solutions of vanishing one-loop Yukawa $\beta$-functions: 
   \beq
   \beta_{ijk}=0.
   \eeq
\end{enumerate}
Then, each of the solutions (\ref{soltheo}) can be uniquely extended
to a formal power series in $g$, and the associated super Yang-Mills
models depend on the single coupling constant $g$ with a $\beta$
function which vanishes at all-orders.

\bigskip

It is important to note a few things:
The requirement of isolated and non-degenerate
solutions guarantees the 
existence of a unique formal power series solution to the reduction
equations.  
The vanishing of the gauge $\beta$-function at one-loop,
$\beta_g^{(1)}$, is equivalent to the 
vanishing of the R current anomaly (\ref{anomaly}).  The vanishing of
the anomalous 
dimensions at one-loop implies the vanishing of the Yukawa couplings
$\beta$-functions at that order.  It also implies the vanishing of the
chiral anomaly coefficients $r^A$.  This last property is a necessary
condition for having $\beta$ functions vanishing at all orders
\footnote{There is an alternative way to find finite theories
  \cite{Leigh:1995ep}.}.  

\bigskip

{\bf Proof:}

Insert $\beta_{ijk}$ as given by the REs into the
relationship (\ref{rbeta}) between the axial anomalies coefficients and
the $\beta$-functions.  Since these chiral anomalies vanish, we get
for $\beta_g$ an homogeneous equation of the form
\beq
0=\beta_g(1+O(\hbar)).
\label{prooftheo}
\eeq
The solution of this equation in the sense of a formal power series in
$\hbar$ is $\beta_g=0$, order by order.  Therefore, due to the
REs (\ref{redeq2}), $\beta_{ijk}=0$ too.

Thus we see that finiteness and reduction of couplings are intimately
related. Since an equation like eq.~(\ref{rbeta}) is lacking in
non-supersymmetric theories, one cannot extend the validity of a
similar theorem in such theories.

\section{Sum rule for SB terms in \boldmath$N=1$ Supersymmetric and Finite theories: All-loop results}

The method of reducing the dimensionless couplings has been
extended\cite{Kubo:1996js}, to the soft SUSY breaking (SSB)
dimensionful parameters of $N = 1$ supersymmetric theories.  In
addition it was found \cite{Kawamura:1997cw} that RGI SSB scalar
masses in Gauge-Yukawa unified models satisfy a universal sum rule.
Here we will describe first how the use of the available two-loop RG
functions and the requirement of finiteness of the SSB parameters up
to this order leads to the soft scalar-mass sum rule
\cite{Kobayashi:1997qx}.

Consider the superpotential given by (\ref{supot}) 
along with the Lagrangian for SSB terms
\bea
-{\cal L}_{\rm SB} &=&
\frac{1}{6} \,h^{ijk}\,\phi_i \phi_j \phi_k
+
\frac{1}{2} \,b^{ij}\,\phi_i \phi_j \nn\\
&+&
\frac{1}{2} \,(m^2)^{j}_{i}\,\phi^{*\,i} \phi_j+
\frac{1}{2} \,M\,\lambda \lambda+\mbox{h.c.},
\eea
where the $\phi_i$ are the
scalar parts of the chiral superfields $\Phi_i$ , $\lambda$ are the gauginos
and $M$ their unified mass.
Since we would like to consider
only finite theories here, we assume that 
the gauge group is  a simple group and the one-loop
$\beta$-function of the 
gauge coupling $g$  vanishes.
We also assume that the reduction equations 
admit power series solutions of the form
\bea 
C^{ijk} &=& g\,\sum_{n}\,\rho^{ijk}_{(n)} g^{2n}~.
\label{Yg}
\eea 
According to the finiteness theorem
of ref.~\cite{Lucchesi:1987ef,Lucchesi:1987he}, the theory is then finite to all orders in
perturbation theory, if, among others, the one-loop anomalous dimensions
$\gamma_{i}^{j(1)}$ vanish.  The one- and two-loop finiteness for
$h^{ijk}$ can be achieved by \cite{Jack:1994kd}
\bea h^{ijk} &=& -M C^{ijk}+\dots =-M
\rho^{ijk}_{(0)}\,g+O(g^5)~,
\label{hY}
\eea
where $\dots$ stand for  higher order terms. 

Now, to obtain the two-loop sum rule for 
soft scalar masses, we assume that 
the lowest order coefficients $\rho^{ijk}_{(0)}$ 
and also $(m^2)^{i}_{j}$ satisfy the diagonality relations
\bea
\rho_{ipq(0)}\rho^{jpq}_{(0)} &\propto & \delta_{i}^{j}~\mbox{for all} 
~p ~\mbox{and}~q~~\mbox{and}~~
(m^2)^{i}_{j}= m^{2}_{j}\delta^{i}_{j}~,
\label{cond1}
\eea
respectively.
Then we find the following soft scalar-mass sum
rule \cite{Kobayashi:1997qx,Kobayashi:1999pn,Mondragon:2003bp}
\bea
(~m_{i}^{2}+m_{j}^{2}+m_{k}^{2}~)/
M M^{\dag} &=&
1+\frac{g^2}{16 \pi^2}\,\Delta^{(2)}+O(g^4)~
\label{sumr} 
\eea
for i, j, k with $\rho^{ijk}_{(0)} \neq 0$, where $\Delta^{(2)}$ is
the two-loop correction
\bea
\Delta^{(2)} &=&  -2\sum_{l} [(m^{2}_{l}/M M^{\dag})-(1/3)]~T(R_l),
\label{delta}
\eea
which vanishes for the
universal choice in accordance with the previous findings of
ref.~\cite{Jack:1994kd}.

If we know higher-loop $\beta$-functions explicitly, we can follow the same 
procedure and find higher-loop RGI relations among SSB terms.
However, the $\beta$-functions of the soft scalar masses are explicitly
known only up to two loops.
In order to obtain higher-loop results some relations among 
$\beta$-functions are needed.

Making use of the spurion technique
\cite{Delbourgo:1974jg,Salam:1974pp,Fujikawa:1974ay,Grisaru:1979wc,Girardello:1981wz}, it is possible to find 
the following  all-loop relations among SSB $\beta$-functions, 
\cite{Hisano:1997ua,Jack:1997pa,Avdeev:1997vx,Kazakov:1998uj,Kazakov:1997nf,Jack:1997eh}
\bea
\beta_M &=& 2{\cal O}\left(\frac{\beta_g}{g}\right)~,
\label{betaM}\\
\beta_h^{ijk}&=&\gamma^i{}_lh^{ljk}+\gamma^j{}_lh^{ilk}
+\gamma^k{}_lh^{ijl}\nn\\
&&-2\gamma_1^i{}_lC^{ljk}
-2\gamma_1^j{}_lC^{ilk}-2\gamma_1^k{}_lC^{ijl}~,\\
(\beta_{m^2})^i{}_j &=&\left[ \Delta 
+ X \frac{\partial}{\partial g}\right]\gamma^i{}_j~,
\label{betam2}\\
{\cal O} &=&\left(Mg^2\frac{\partial}{\partial g^2}
-h^{lmn}\frac{\partial}{\partial C^{lmn}}\right)~,
\label{diffo}\\
\Delta &=& 2{\cal O}{\cal O}^* +2|M|^2 g^2\frac{\partial}
{\partial g^2} +\tilde{C}_{lmn}
\frac{\partial}{\partial C_{lmn}} +
\tilde{C}^{lmn}\frac{\partial}{\partial C^{lmn}}~,
\eea
where $(\gamma_1)^i{}_j={\cal O}\gamma^i{}_j$, 
$C_{lmn} = (C^{lmn})^*$, and 
\bea
\tilde{C}^{ijk}&=&
(m^2)^i{}_lC^{ljk}+(m^2)^j{}_lC^{ilk}+(m^2)^k{}_lC^{ijl}~.
\label{tildeC}
\eea
It was also found \cite{Jack:1997pa}  that the relation
\bea
h^{ijk} &=& -M (C^{ijk})'
\equiv -M \frac{d C^{ijk}(g)}{d \ln g}~,
\label{h2}
\eea
among couplings is all-loop RGI. Furthermore, using the all-loop gauge
$\beta$-function of Novikov {\em et al.} 
\cite{Novikov:1983ee,Novikov:1985rd,Shifman:1996iy} given
by 
\bea
\beta_g^{\rm NSVZ} &=& 
\frac{g^3}{16\pi^2} 
\left[ \frac{\sum_l T(R_l)(1-\gamma_l /2)
-3 C(G)}{ 1-g^2C(G)/8\pi^2}\right]~, 
\label{bnsvz}
\eea 
it was found the all-loop RGI sum rule \cite{Kobayashi:1998jq},
\bea
m^2_i+m^2_j+m^2_k &=&
|M|^2 \{~
\frac{1}{1-g^2 C(G)/(8\pi^2)}\frac{d \ln C^{ijk}}{d \ln g}
+\frac{1}{2}\frac{d^2 \ln C^{ijk}}{d (\ln g)^2}~\}\nn\\
& &+\sum_l
\frac{m^2_l T(R_l)}{C(G)-8\pi^2/g^2}
\frac{d \ln C^{ijk}}{d \ln g}~.
\label{sum2}
\eea
In addition 
the exact-$\beta$-function for $m^2$
in the NSVZ scheme has been obtained \cite{Kobayashi:1998jq} for the first time and
is given by
\bea
\beta_{m^2_i}^{\rm NSVZ} &=&\left[~
|M|^2 \{~
\frac{1}{1-g^2 C(G)/(8\pi^2)}\frac{d }{d \ln g}
+\frac{1}{2}\frac{d^2 }{d (\ln g)^2}~\}\right.\nn\\
& &\left. +\sum_l
\frac{m^2_l T(R_l)}{C(G)-8\pi^2/g^2}
\frac{d }{d \ln g}~\right]~\gamma_{i}^{\rm NSVZ}~.
\label{bm23}
\eea
Surprisingly enough, the all-loop result (\ref{sum2}) coincides with 
the superstring result for the finite case in a certain class of 
orbifold models \cite{Kobayashi:1997qx} if \\ 
$d \ln C^{ijk}/{d \ln g}=1$.


\section{Finite \boldmath${SU(5)}$ Unified Theories}

Finite Unified Theories (FUTs) have always attracted interest for
their intriguing mathematical properties and their predictive power.
One very important result is that the one-loop finiteness conditions
(\ref{betay},\ref{gamay}) are sufficient to guarantee two-loop
finiteness \cite{Parkes:1984dh}.  A classification of possible
one-loop finite models was done by two groups
\cite{Hamidi:1984ft,Jiang:1988na,Jiang:1987hv}.  The first one and
two-loop finite $SU(5)$ model was presented in \cite{Jones:1984qd},
and shortly afterwards the conditions for finiteness in the soft
SUSY-breaking sector at one-loop \cite{Jones:1984cx} were given.  In
\cite{Leon:1985jm} a one and two-loop finite $SU(5)$ model was
presented where the rotation of the Higgs sector was proposed as a way
of making it realistic.  The first all-loop finite theory was studied
in \cite{Kapetanakis:1992vx,Mondragon:1993tw}, without taking into
account the soft breaking terms. Finite soft breaking terms and the
proof that one-loop finiteness in the soft terms also implies two-loop
finiteness was done in \cite{Jack:1994kd}.  The inclusion of soft
breaking terms in a realistic model was done in \cite{Kazakov:1995cy}
and their finiteness to all-loops studied in \cite{Kazakov:1997nf},
although the universality of the soft breaking terms lead to a charged
LSP. This fact was also noticed in \cite{Yoshioka:1997yt}, where the
inclusion of an extra parameter in the boundary condition of the Higgs
mixing mass parameter was introduced to alleviate
it. 
The derivation of the sum-rule in the soft SUSY breaking
sector and the proof that it can be made all-loop finite were done in
refs.~\cite{Kobayashi:1997qx} and \cite{Kobayashi:1998jq}
respectively, allowing thus for the construction of all-loop finite
realistic models.

From the classification of theories with vanishing one-loop gauge
$\beta$-function \cite{Hamidi:1984ft}, one can easily see that there
exist only two candidate possibilities to construct $SU(5)$ GUTs with
three generations. These possibilities require that the theory should
contain as matter fields the chiral supermultiplets ${\bf
  5},~\overline{\bf 5},~{\bf 10}, ~\overline{\bf 5},~{\bf 24}$ with
the multiplicities $(6,9,4,1,0)$ or $(4,7,3,0,1)$, respectively. Only
the second one contains a ${\bf 24}$-plet which can be used to provide
the spontaneous symmetry breaking (SB) of $SU(5)$ down to $SU(3)\times
SU(2)\times U(1)$. For the first model one has to incorporate another
way, such as the Wilson flux breaking mechanism to achieve the desired
SB of $SU(5)$ \cite{Kapetanakis:1992vx,Mondragon:1993tw}. Therefore,
for a self-consistent field theory discussion we would like to
concentrate only on the second possibility.

The particle content of the models we will study consists of the
following supermultiplets: three ($\overline{\bf 5} + \bf{10}$),
needed for each of the three generations of quarks and leptons, four
($\overline{\bf 5} + {\bf 5}$) and one ${\bf 24}$ considered as Higgs
supermultiplets. 
When the gauge group of the finite GUT is broken the theory is no
longer finite, and we will assume that we are left with the MSSM.

Therefore, a predictive Gauge-Yukawa unified $SU(5)$
model which is finite to all orders, in addition to the requirements
mentioned already, should also have the following properties:

\begin{enumerate}

\item 
One-loop anomalous dimensions are diagonal,
i.e.,  $\gamma_{i}^{(1)\,j} \propto \delta^{j}_{i} $.
\item The three fermion generations, in the irreducible representations
  $\overline{\bf 5}_{i},{\bf 10}_i$ $(i=1,2,3)$,  should
  not couple to the adjoint ${\bf 24}$.
\item The two Higgs doublets of the MSSM should mostly be made out of a
pair of Higgs quintet and anti-quintet, which couple to the third
generation.
\end{enumerate}

In the following we discuss two versions of the all-order finite
model:  the model of \citere{Kapetanakis:1992vx,Mondragon:1993tw},
which will be labeled ${\bf A}$, and a slight variation of this model
(labeled ${\bf B}$), which can also be obtained from the class of the
models suggested in \citere{Avdeev:1997vx,Kazakov:1998uj} with a
modification to suppress non-diagonal anomalous dimensions
\cite{Kobayashi:1997qx}.

The superpotential which describes the two models before the reduction
of couplings takes places is of the form
\cite{Kapetanakis:1992vx,Mondragon:1993tw,Kobayashi:1997qx,Jones:1984qd,Leon:1985jm} 
\bea 
W &=&
\sum_{i=1}^{3}\,[~\frac{1}{2}g_{i}^{u} \,{\bf 10}_i{\bf 10}_i H_{i}+
g_{i}^{d}\,{\bf 10}_i \overline{\bf 5}_{i}\, \overline{H}_{i}~] +
g_{23}^{u}\,{\bf 10}_2{\bf 10}_3 H_{4} \label{zoup-super1}\\
& &+g_{23}^{d}\,{\bf 10}_2 \overline{\bf 5}_{3}\, \overline{H}_{4}+
g_{32}^{d}\,{\bf 10}_3 \overline{\bf 5}_{2}\, \overline{H}_{4}+
\sum_{a=1}^{4}g_{a}^{f}\,H_{a}\, {\bf 24}\,\overline{H}_{a}+
\frac{g^{\lambda}}{3}\,({\bf 24})^3~,\nonumber
\eea
where 
$H_{a}$ and $\overline{H}_{a}~~(a=1,\dots,4)$
stand for the Higgs quintets and anti-quintets.

 The main difference between model ${\bf A}$ and model
${\bf B}$ is that two pairs of Higgs quintets and anti-quintets couple
to the ${\bf 24}$ in ${\bf B}$, so that it is not necessary to mix
them with $H_{4}$ and $\overline{H}_{4}$ in order to achieve the
triplet-doublet splitting after the symmetry breaking of $SU(5)$
\cite{Kobayashi:1997qx}.  Thus, although the particle content is the
same, the solutions to Eqs.~(\ref{betay},\ref{gamay}) and the sum rules
are different, which will reflect in the phenomenology, as we will
see.

\subsection{\FUTA }
\begin{table}
\begin{center}
\renewcommand{\arraystretch}{1.3}
\begin{tabular}{|l|l|l|l|l|l|l|l|l|l|l|l|l|l|l|l|}
\hline
& $\overline{{\bf 5}}_{1} $ & $\overline{{\bf 5}}_{2} $& $\overline{{\bf
    5}}_{3}$ & ${\bf 10}_{1} $ &  ${\bf 10}_{2}$ & ${\bf
  10}_{3} $ & $H_{1} $ & $H_{2} $ & $H_{3} $ &$H_{4 }$&  $\overline H_{1} $ &
$\overline H_{2} $ & $\overline H_{3} $ &$\overline H_{4 }$& ${\bf 24} $\\ \hline
$Z_7$ & 4 & 1 & 2 & 1 & 2 & 4 & 5 & 3 & 6 & -5 & -3 & -6 &0& 0 & 0 \\\hline
$Z_3$ & 0 & 0 & 0 & 1 & 2 & 0 & 1 & 2 & 0 & -1 & -2 & 0 & 0 & 0&0  \\\hline
$Z_2$ & 1 & 1 & 1 & 1 & 1 & 1 & 0 & 0 & 0 & 0 & 0 & 0 &  0 & 0 &0 \\\hline
\end{tabular}
  \caption{Charges of the $Z_7\times Z_3\times Z_2$ symmetry for Model
    \FUTA. }
\renewcommand{\arraystretch}{1.0}
\label{tableA}
\end{center}
\end{table}

After the reduction of couplings 
the symmetry of the superpotential $W$ (\ref{zoup-super1}) is enhanced.
For  model ${\bf A}$ one finds that
the superpotential has the
$Z_7\times Z_3\times Z_2$ discrete symmetry with the charge assignment
as shown in Table \ref{tableA}, and with the following superpotential
\beq
W_A = \sum_{i=1}^{3}\,[~\frac{1}{2}g_{i}^{u}
\,{\bf 10}_i{\bf 10}_i H_{i}+
g_{i}^{d}\,{\bf 10}_i \overline{\bf 5}_{i}\,
\overline{H}_{i}~] +
g_{4}^{f}\,H_{4}\, 
{\bf 24}\,\overline{H}_{4}+
\frac{g^{\lambda}}{3}\,({\bf 24})^3~,
\label{w-futa}
\eeq

The non-degenerate and isolated solutions to $\gamma^{(1)}_{i}=0$ for
 model \FUTA, which are the boundary conditions for the Yukawa
 couplings at the GUT scale, are: 
\bea 
&& (g_{1}^{u})^2
=\frac{8}{5}~g^2~, ~(g_{1}^{d})^2
=\frac{6}{5}~g^2~,~
(g_{2}^{u})^2=(g_{3}^{u})^2=\frac{8}{5}~g^2~,\label{zoup-SOL5}\\
&& (g^{\lambda})^2 =\frac{15}{7}g^2~,~
 (g_{2}^{d})^2 = (g_{3}^{d})^2=\frac{6}{5}~g^2~,~(g_{4}^{f})^2= g^2\nonumber\\
&&(g_{23}^{u})^2 = 
(g_{23}^{d})^2=(g_{32}^{d})^2=
 (g_{2}^{f})^2=
(g_{3}^{f})^2=(g_{1}^{f})^2=0~.~\nonumber 
\eea 
In the dimensionful sector, the sum rule gives us the following
boundary conditions at the GUT scale for this model
\cite{Kobayashi:1997qx}: 
\bea
m^{2}_{H_u}+
2  m^{2}_{{\bf 10}} &=&
m^{2}_{H_d}+ m^{2}_{\overline{{\bf 5}}}+
m^{2}_{{\bf 10}}=M^2 ~~,
\eea
and thus we are left with only three free parameters, namely
$m_{\overline{{\bf 5}}}\equiv m_{\overline{{\bf 5}}_3}$, 
$m_{{\bf 10}}\equiv m_{{\bf 10}_3}$
and $M$.

\subsection{\FUTB}
Also in the case of \FUTB\ the symmetry is enhanced after the reduction
of couplings.  The superpotential has now a 
  $Z_4\times Z_4\times Z_4$ symmetry with charges as shown in Table
\ref{tableB} and  with the
following superpotential
\begin{multline}
W_B = \sum_{i=1}^{3}\,[~\frac{1}{2}g_{i}^{u}
\,{\bf 10}_i{\bf 10}_i H_{i}+
g_{i}^{d}\,{\bf 10}_i \overline{\bf 5}_{i}\,
\overline{H}_{i}~] +
g_{23}^{u}\,{\bf 10}_2{\bf 10}_3 H_{4} \\
  +g_{23}^{d}\,{\bf 10}_2 \overline{\bf 5}_{3}\,
\overline{H}_{4}+
g_{32}^{d}\,{\bf 10}_3 \overline{\bf 5}_{2}\,
\overline{H}_{4}+
g_{2}^{f}\,H_{2}\, 
{\bf 24}\,\overline{H}_{2}+ g_{3}^{f}\,H_{3}\, 
{\bf 24}\,\overline{H}_{3}+
\frac{g^{\lambda}}{3}\,({\bf 24})^3~,
\label{w-futb}
\end{multline}
For this model the non-degenerate and isolated solutions to
$\gamma^{(1)}_{i}=0$ give us: 
\bea 
&& (g_{1}^{u})^2
=\frac{8}{5}~ g^2~, ~(g_{1}^{d})^2
=\frac{6}{5}~g^2~,~
(g_{2}^{u})^2=(g_{3}^{u})^2=(g_{23}^{u})^2 =\frac{4}{5}~g^2~,\label{zoup-SOL52}
\nonumber\\
&& (g_{2}^{d})^2 = (g_{3}^{d})^2=
(g_{23}^{d})^2=(g_{32}^{d})^2=\frac{3}{5}~g^2~,
\\
&& (g^{\lambda})^2 =\frac{15}{7}g^2~,~ (g_{2}^{f})^2
=(g_{3}^{f})^2=\frac{1}{2}~g^2~,~ (g_{1}^{f})^2=
(g_{4}^{f})^2=0~,\nonumber 
\eea 
and from the sum rule we obtain \cite{Kobayashi:1997qx}:
\bea
m^{2}_{H_u}+
2  m^{2}_{{\bf 10}} &=&M^2~,~
m^{2}_{H_d}-2m^{2}_{{\bf 10}}=-\frac{M^2}{3}~,~\nonumber\\
m^{2}_{\overline{{\bf 5}}}+
3m^{2}_{{\bf 10}}&=&\frac{4M^2}{3}~,
\eea
i.e., in this case we have only two free parameters  
$m_{{\bf 10}}\equiv m_{{\bf 10}_3}$  and $M$ for the dimensionful sector.

\begin{table}
\begin{center}
\renewcommand{\arraystretch}{1.3}
\begin{tabular}{|l|l|l|l|l|l|l|l|l|l|l|l|l|l|l|l|}
\hline
& $\overline{{\bf 5}}_{1} $ & $\overline{{\bf 5}}_{2} $& $\overline{{\bf
    5}}_{3}$ & ${\bf 10}_{1} $ &  ${\bf 10}_{2}$ &  ${\bf
  10}_{3} $ & $ H_{1} $ & $H_{2} $ & $ H_{3}
$ &$H_{4}$&   $\overline H_{1} $ & 
$\overline H_{2} $ & $\overline H_{3} $ &$\overline H_{4 }$&${\bf 24} $\\\hline
$Z_4$ & 1 & 0 & 0 & 1 & 0 & 0 & 2 & 0 & 0 & 0 & -2 & 0 & 0 & 0 &0  \\\hline
$Z_4$ & 0 & 1 & 0 & 0 & 1 & 0 & 0 & 2 & 0 & 3 & 0 & -2 & 0 & -3& 0  \\\hline
$Z_4$ & 0 & 0 & 1 & 0 & 0 & 1 & 0 & 0 & 2 & 3 & 0 & 0 & -2& -3 & 0 \\\hline
\end{tabular}
  \caption{Charges of the $Z_4\times Z_4\times Z_4$ symmetry for Model
    \FUTB.}
\label{tableB}
\renewcommand{\arraystretch}{1.0}
\end{center}

\end{table}

As already mentioned, after the $SU(5)$ gauge symmetry breaking we
assume we have the MSSM, i.e. only two Higgs doublets.  This can be
achieved by introducing appropriate mass terms that allow to perform a
rotation of the Higgs sector \cite{Leon:1985jm, Kapetanakis:1992vx,Mondragon:1993tw,Hamidi:1984gd,
  Jones:1984qd}, in such a way that only one pair of Higgs doublets,
coupled mostly to the third family, remains light and acquire vacuum
expectation values.  To avoid fast proton decay the usual fine tuning
to achieve doublet-triplet splitting is performed.  Notice that,
although similar, the mechanism is not identical to minimal $SU(5)$,
since we have an extended Higgs sector.

Thus, after the gauge symmetry of the GUT theory is broken we are left
with the MSSM, with the boundary conditions for the third family given
by the finiteness conditions, while the other two families are basically
decoupled.

We will now examine the phenomenology of such all-loop Finite Unified
theories with $SU(5)$ gauge group and, for the reasons expressed
above,   we will concentrate only on the
third generation of quarks and leptons. An extension to three
families, and the generation of quark mixing angles and masses in
Finite Unified Theories has been addressed in \cite{Babu:2002in},
where several examples are given. These extensions are not considered
here.  


\subsection{Restrictions from low-energy observables}
\label{sec:ewpo}

Since the gauge symmetry is spontaneously broken below $M_{\rm GUT}$,
the finiteness conditions do not restrict the renormalization
properties at low energies, and all it remains are boundary conditions
on the gauge and Yukawa couplings (\ref{zoup-SOL5}) or
(\ref{zoup-SOL52}), the $h=-MC$ relation (\ref{hY}), and the soft
scalar-mass sum rule (\ref{sumr}) at $M_{\rm GUT}$, as applied in
the two models.  Thus we examine the evolution of these parameters
according to their RGEs up to two-loops for dimensionless parameters
and at one-loop for dimensionful ones with the relevant boundary
conditions.  Below $M_{\rm GUT}$ their evolution is assumed to be
governed by the MSSM.  We further assume a unique SUSY
breaking scale $M_{\rm SUSY}$ (which we define as the geometrical average
of the stop masses) and therefore below that scale the effective
theory is just the SM.  This allows to evaluate observables at or
below the electroweak scale.

In the following, we briefly describe the low-energy observables used in
our analysis. We discuss the current precision of
the experimental results and the theoretical predictions. 
We also give relevant details of the higher-order perturbative
corrections that we include. 
We do not discuss theoretical
uncertainties from the RG running between the high-scale parameters
and the weak scale.
At present, these uncertainties are expected to be 
less important than the experimental and theoretical uncertainties of
the precision observables. 

As precision observables we first discuss the 3rd generation quark
masses that are leading to the strongest constraints on the models under
investigation. Next we apply $B$~physics and Higgs-boson mass
constraints.  We also briefly discuss the anomalous magnetic moment of the
muon.


\subsection{Predictions}
\begin{figure}[htp]
\vspace{0.5cm}
           \centerline{\includegraphics[width=10cm,angle=0]{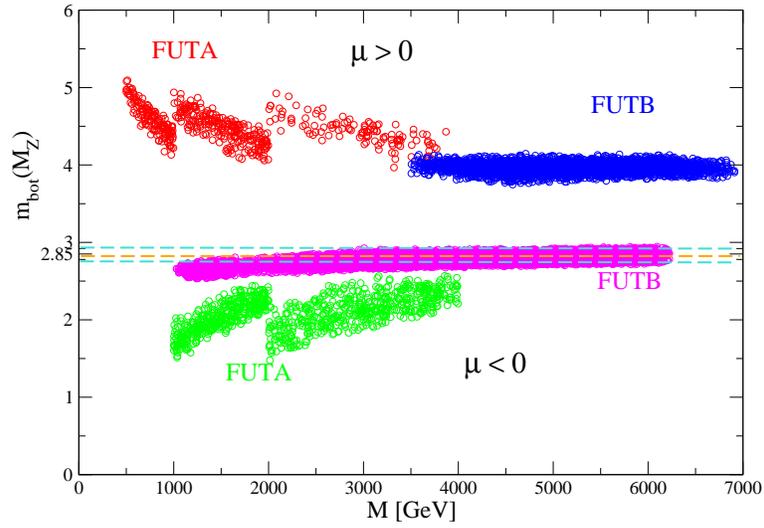}}
\vspace{2cm}
           \centerline{\includegraphics[width=10cm,angle=0]{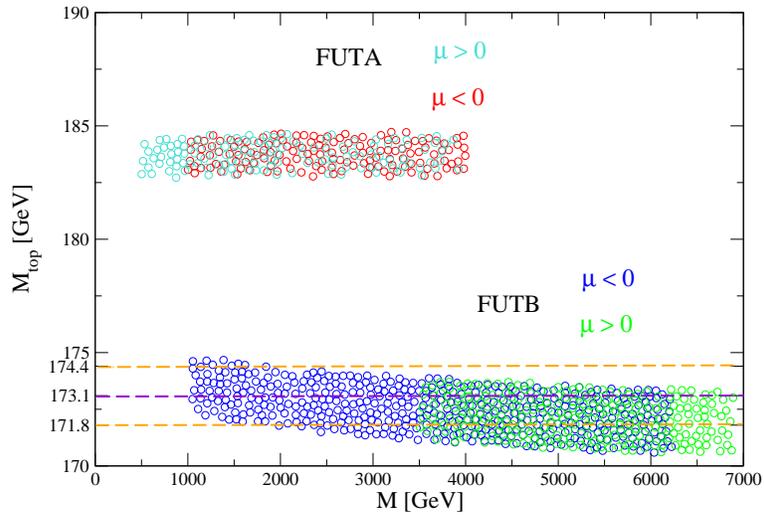}}
\vspace{0.5cm}
       \caption{The bottom quark mass at the $Z$~boson scale (upper) 
                and top quark pole mass (lower plot) are shown 
                as function of $M$ for both models.}
\label{fig:MtopbotvsM}
\vspace{-0.5em}
\end{figure}

We now present the comparison of the predictions of the four models with
the experimental data, see ref.~\cite{Heinemeyer:2007tz} for more details,
starting with the heavy quark masses. 
In fig.\ref{fig:MtopbotvsM} we show the {\bf FUTA} and {\bf FUTB}
predictions for the top pole mass, $M_{\rm top}$, and the running bottom
mass at the scale $\MZ$, $m_{\rm bot}(M_Z)$, 
as a function of the  unified gaugino mass $M$, for the two cases
$\mu <0$ and $\mu >0$.
The running bottom mass is used to avoid the large
QCD uncertainties inherent for the pole mass. 
In the evaluation of the bottom mass $m_{\rm bot}$,
we have included the corrections coming from bottom
squark-gluino loops and top squark-chargino loops~\cite{Carena:1999py}.  We
compare the predictions for the running bottom quark mass with the
experimental value, 
$m_{b}(M_Z) = 2.83 \pm 0.10 \gev$~\cite{Amsler:2008zzb}. One can
see that the value of $m_{\rm bot}$ depends
strongly on the sign of $\mu$ due to the above mentioned 
radiative corrections involving SUSY particles.  
For both models ${\bf A}$ and ${\bf B}$ the values
for $\mu >0$ are above the central experimental value, with 
$m_{\rm bot}(M_Z) \sim 4.0 - 5.0$~GeV.  
For $\mu < 0$, on the other hand, model ${\bf B}$ shows
overlap with the experimentally measured values, 
$m_{\rm bot}(M_Z) \sim 2.5-2.8$~GeV.
For model ${\bf A}$ we find $m_{\rm bot}(M_Z) \sim 1.5 - 2.6$~GeV, and 
there is only a small region of allowed parameter space at large $M$
where we find agreement with the experimental value at the two~$\si$ level.
In summary, the experimental determination of $m_{\rm bot}(\MZ$) clearly
selects the negative sign of $\mu$. 

Now we turn to the top quark mass.  The predictions for the top quark
mass $M_{\rm top}$ are $\sim 183$ and $\sim 172$~GeV in the models
${\bf A}$ and ${\bf B}$ respectively, as shown in the lower plot of
fig.~\ref{fig:MtopbotvsM}.  Comparing these predictions with
the experimental value $m_{t}^{exp} = (173.1 \pm
1.3)$~GeV~\cite{:2009ec},\footnote{Using the most recent value of
  $m_t^{exp} = 173.3 \pm 1.1 \gev$~\cite{:1900yx} does not
  significantly change our results.} and recalling that the
theoretical values for $M_{\rm top}$ may suffer from a correction of
$\sim 4 \%$ \cite{Kubo:1997fi,Kobayashi:2001me,Mondragon:2003bp}, we
see that clearly model ${\bf B}$ is singled out.  In addition the
value of $\tan \beta$ is found to be $\tan \beta \sim 54$ and $\sim
48$ for models ${\bf A}$ and ${\bf B}$, respectively.  Thus from the
comparison of the predictions of the two models with experimental data
only {\bf FUTB} with $\mu < 0$ survives.

We now analyze the impact of further low-energy observables on the model
{\bf FUTB} with $\mu < 0$. 
As  additional constraints we consider the following observables: 
the rare $b$~decays $\br(b \to s \gamma)$ and $\br(B_s \to \mu^+ \mu^-)$, 
the lightest Higgs boson mass 
as well as the density of cold dark matter in the Universe, assuming it
consists mainly of neutralinos. More details and a complete set of
references can be found in ref.~\cite{Heinemeyer:2007tz}.

For the branching ratio $\br(b \to s \gamma)$, we take
the experimental value estimated by the Heavy Flavour Averaging
Group (HFAG) is~\cite{Barate:1998vz,Chen:2001fja,Koppenburg:2004fz} 
\beq 
\br(b \to s \gamma ) = (3.55 \pm 0.24 {}^{+0.09}_{-0.10} \pm 0.03) 
                       × 10^{-4},
\label{bsgaexp}
\eeq
where the first error is the combined statistical and uncorrelated systematic 
uncertainty, the latter two errors are correlated systematic theoretical
uncertainties and corrections respectively. 

For the branching ratio $\br(B_s \to \mu^+ \mu^-)$, the SM prediction is
at the level of $10^{-9}$, while the present
experimental upper limit from the Tevatron is 
$4.7 ~ 10^{-8}$ at the $95\%$ C.L.~\cite{hfag}, still providing the
possibility for the MSSM to dominate the SM contribution.

Concerning the lightest Higgs boson mass, $M_h$, the SM bound of
$114.4$ GeV \cite{Barate:2003sz,Schael:2006cr} can be applied, since the
main SM search channels are not suppressed in {\rm FUTB}. For the
prediction we use the code 
{\tt FeynHiggs}
\cite{Heinemeyer:1998yj,Heinemeyer:1998np,Degrassi:2002fi,Frank:2006yh}.  
\begin{figure}[htp]
           \centerline{\includegraphics[width=12cm,angle=0]{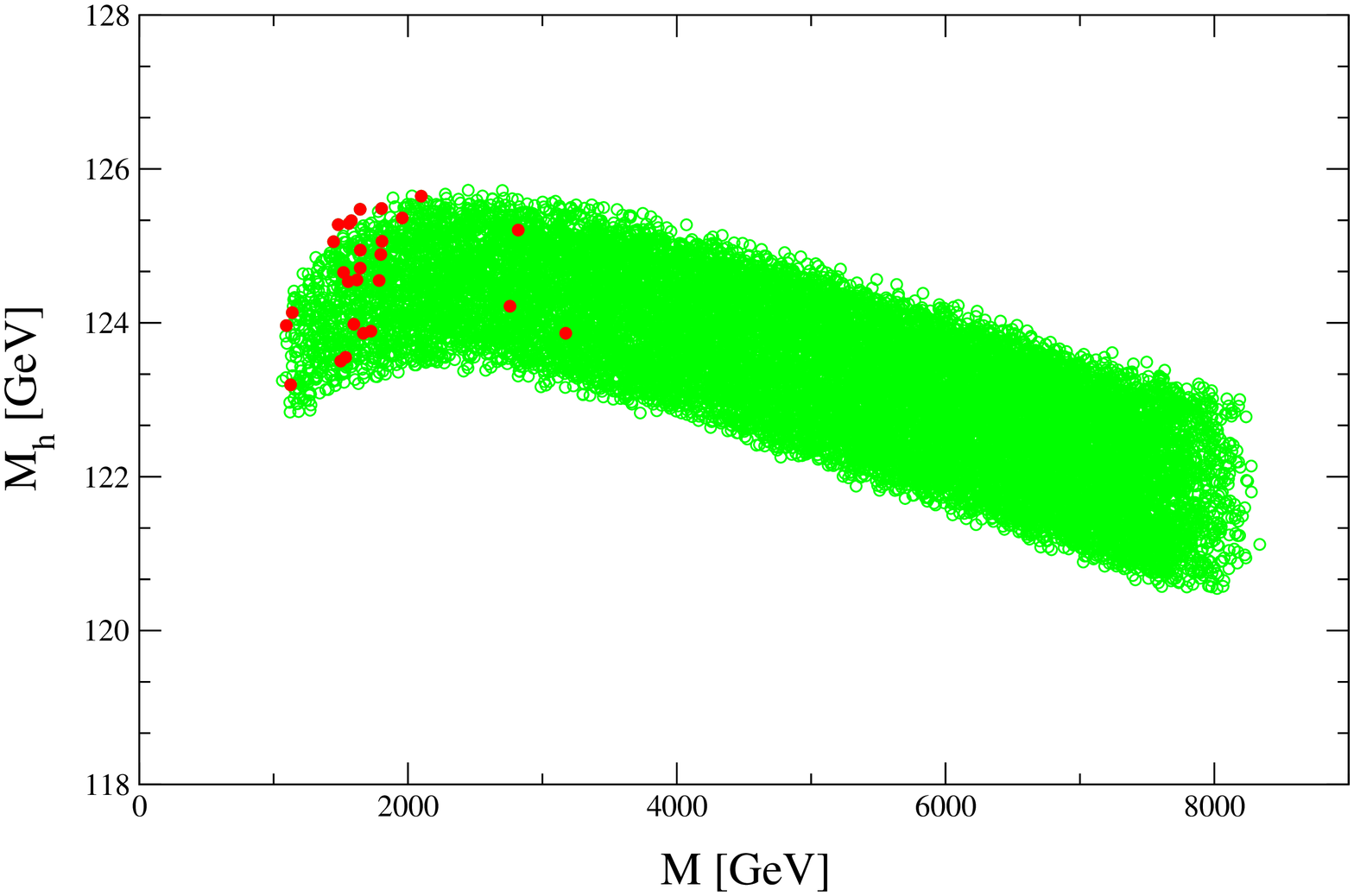}}
        \caption{The lightest Higgs mass, $M_h$,  as function of $M$ for
          the model {\bf FUTB} with $\mu < 0$, see text.}
\label{fig:Higgs}
\end{figure}

The prediction of the lightest Higgs boson mass as a function of $M$ is
shown in \reffi{fig:Higgs}. The light (green) points shown are in agreement
with the two $B$-physics observables listed above. 
The lightest Higgs mass ranges in 
\beq
M_h \sim 121-126~{\rm GeV} , 
\label{eq:Mhpred}
\eeq 
where the uncertainty comes from
variations of the soft scalar masses, and
from finite (i.e.~not logarithmically divergent) corrections in
changing renormalization scheme.  To this value one has to add $\pm 3$
GeV coming from unknown higher order corrections~\cite{Degrassi:2002fi}. 
We have also included a small variation,
due to threshold corrections at the GUT scale, of up to $5 \%$ of the
FUT boundary conditions.  
Thus, taking into account the $B$~physics constraints 
 results naturally in a light Higgs boson that fulfills
the LEP bounds~\cite{Barate:2003sz,Schael:2006cr}. 

In the same way the whole SUSY particle spectrum can be derived. 
The resulting SUSY masses for {\bf FUTB} with $\mu < 0$ are rather large.
The lightest SUSY particle starts around 500 GeV, with the rest
of the spectrum being very heavy. The observation of SUSY particles at
the LHC or the ILC will only be possible in very favorable parts of the
parameter space. For most parameter combination only a SM-like light
Higgs boson in the range of eq.~(\ref{eq:Mhpred}) can be observed.

We note that with such a heavy SUSY spectrum 
the anomalous magnetic moment of the muon, \mbox{$(g-2)_\mu$}
(with $\amu \equiv (g-2)_\mu/2$), gives only a negligible correction to
the SM prediction.  
The comparison of the experimental result and the SM value (based on the
latest combination using $e^+e^-$ data)~\cite{Davier:2010nc}
\beq
\amuexp-\amutheo = (28.7 \pm 8.0) × 10^{-10}.
\label{delamu}
\eeq
would disfavor {\bf FUTB} with $\mu < 0$ by about $3\,\sigma$. However, 
since the SM is not regarded as excluded by $(g-2)_\mu$, 
we still see {\bf FUTB} with $\mu < 0$ as the only surviving model.

Further restrictions on the parameter space can arise from the
requirement that the lightest SUSY particle (LSP) should give the
right amount of cold dark matter (CDM) abundance. The LSP should be
color neutral, and the lightest neutralino appears to be a suitable
candidate~\cite{Goldberg:1983nd,Ellis:1983ew}.
In the case where
all the soft scalar masses are universal at the unfication scale,
there is no region of $M$ below ${\cal O}$(few~TeV) in which 
$m_{\tilde \tau} > m_{\chi^0}$ is satisfied, 
where $m_{\tilde \tau}$ is the lightest $\tilde \tau$ mass, 
and $m_{\chi^0}$ the lightest neutralino mass. An electrically charged
LSP, however, is not in agreement with CDM searches. But once the
universality condition is relaxed this problem can be solved
naturally, thanks to the sum rule (\ref{sumr}).  Using this equation
a comfortable parameter space is found for 
{\bf FUTB} with $\mu < 0$ (and also for {\bf FUTA} and both signs of $\mu$).
that fulfills the conditions of (a) successful radiative
electroweak symmetry breaking, (b) $m_{\tilde\tau}> m_{\chi^0}$. 

Calculating the CDM abundance in these FUT models one finds that usually
it is very large, thus a
mechanism is needed in our model to reduce it. This issue could, for
instance, be related to another problem, that of neutrino masses.
This type of masses cannot be generated naturally within the class of
finite unified theories that we are considering in this paper,
although a non-zero value for neutrino masses has clearly been
established~\cite{Amsler:2008zzb}.  However, the class of FUTs
discussed here can, in principle, be easily extended by introducing
bilinear R-parity violating terms that preserve finiteness and
introduce the desired neutrino masses \cite{Valle:1998bs}.  R-parity
violation~\cite{Dreiner:1997uz,Bhattacharyya:1997vv,Allanach:1999ic,Romao:1991ex}
would have a small impact on the collider phenomenology discussed here
(apart from fact the SUSY search strategies could not rely on a
``missing energy'' signature), but remove the CDM bound completely.  The
details of such a possibility in the present framework attempting to
provide the models with realistic neutrino masses will be discussed
elsewhere.  Other mechanisms, not involving R-parity violation (and
keeping the ``missing energy'' signature), that could be invoked if the
amount of CDM appears to be too large, concern the cosmology of the
early universe.  For instance, ``thermal
inflation''~\cite{Lyth:1995ka} or ``late time entropy
injection''~\cite{Gelmini:2006pw} could bring the CDM density into
agreement with the WMAP measurements.  This kind of modifications of
the physics scenario neither concerns the theory basis nor the
collider phenomenology, but could have a strong impact on the CDM
derived bounds.

\begin{figure}[htp]
           \centerline{\includegraphics[width=12cm,angle=0]{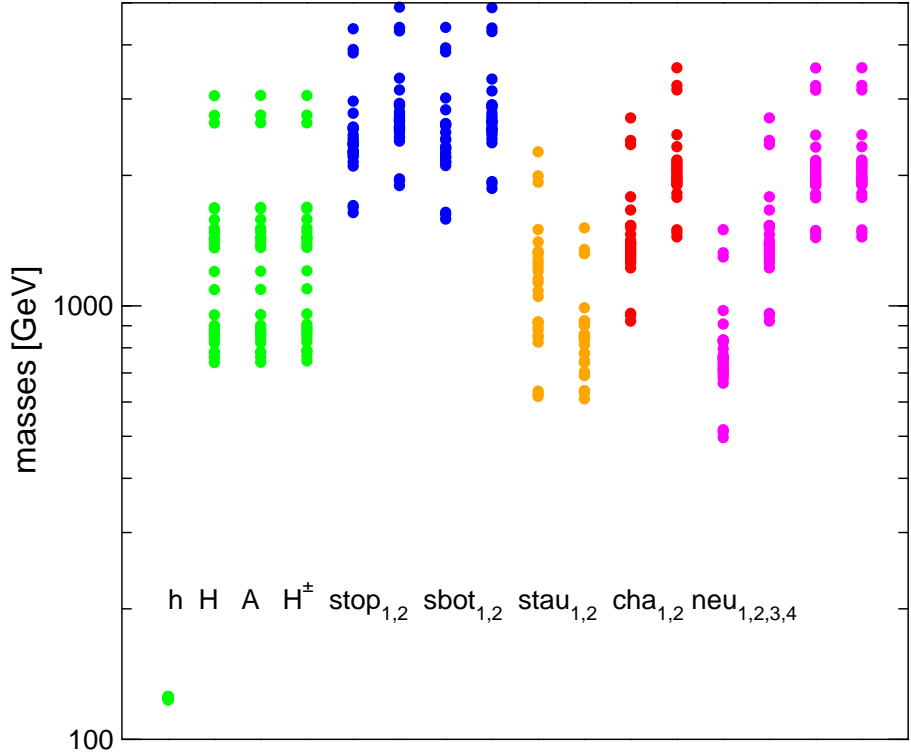}}
           \caption{The particle spectrum of model {\bf FUTB} with $\mu <0$,
             where 
             the points shown are in agreement with the quark mass
             constraints, the  
             $B$-physics observables and the loose CDM constraint.
  The light (green) points on the left are the various Higgs boson masses. The
  dark (blue) points following are the two scalar top and bottom masses,
  followed by lighter (beige) scalar tau masses. The darker (red) points to
  the right are the two chargino masses followed by the lighter shaded (pink)
  points indicating the neutralino masses.}
\label{fig:masses}
\vspace{-0.5em}
\end{figure}

Therefore, in order to get an impression of the
{\em possible} impact of the CDM abundance on the collider phenomenology
in our models under investigation, we will analyze the case that the LSP
does contribute to the CDM density, and apply a more loose bound of  
\BE
\Omega_{\rm CDM} h^2 < 0.3~.
\label{cdmloose}
\EE
(Lower values than the ones permitted by \refeq{cdmloose} are
naturally allowed if another particle than the lightest neutralino
constitutes CDM.)  For our evaluation we have used the code {\tt
  MicroMegas}~\cite{Belanger:2001fz,Belanger:2004yn}.  
The prediction for the lightest Higgs mass, $M_h$ as
function of $M$ for the model {\bf FUTB} with $\mu < 0$
is shown in \reffi{fig:Higgs}. The dark (red) dots are the points that pass
the constraints in \refeq{cdmloose} (and that have the lightest neutralino as
LSP), which favors relatively light values of $M$.
The full particle 
spectrum of model {\bf FUTB} with $\mu <0$, again compliant with quark mass
constraints, the $B$ physics observables (and with the loose CDM constraint), 
is shown in fig.~\reffi{fig:masses}. The masses of the particles increase with
increasing values of the unified gaugino mass $M$.
One can see that large parts of the spectrum are in the kinematic reach of the
LHC. A numerical example of such a light spectrum is shown in
Table~\ref{table:mass}. The colored part of this spectrum as well as the
lightest Higgs boson should be (relatively easily) accessible at the LHC.

\begin{table}
\begin{center}
\begin{tabular}{|l|l||l|l|}
\hline 
Mbot($M_Z$) &  2.71 GeV &
Mtop &    172.2 GeV\\ \hline
Mh &  123.1 GeV & 
MA &  680 GeV\\ \hline 
MH &  679 GeV& 
MH$^\pm$ &  685 GeV \\ \hline 
 Stop1 &  1876 GeV &
Stop2 &    2146 GeV \\ \hline
Sbot1 &   1849 GeV & 
Sbot2 &    2117 GeV\\ \hline 
Mstau1 &    635 GeV & 
Mstau2 &    867 GeV\\ \hline 
Char1 &    1072 GeV & 
Char2 &    1597 GeV\\ \hline
Neu1  &    579 GeV &
Neu2  &    1072 GeV \\ \hline 
Neu3  &    1591 GeV &
Neu4  &    1596 GeV \\ \hline
 M1 &    580 GeV& 
 M2 &   1077 GeV\\ \hline 
 Mgluino &    2754 GeV & 
&  \\
\hline 
\end{tabular}
\caption{A representative spectrum of a light {\bf FUTB}, $\mu <0$ spectrum.}
\label{table:mass}
\end{center}
\end{table}

\begin{figure}[htp]
           \centerline{\includegraphics[width=12cm,angle=0]{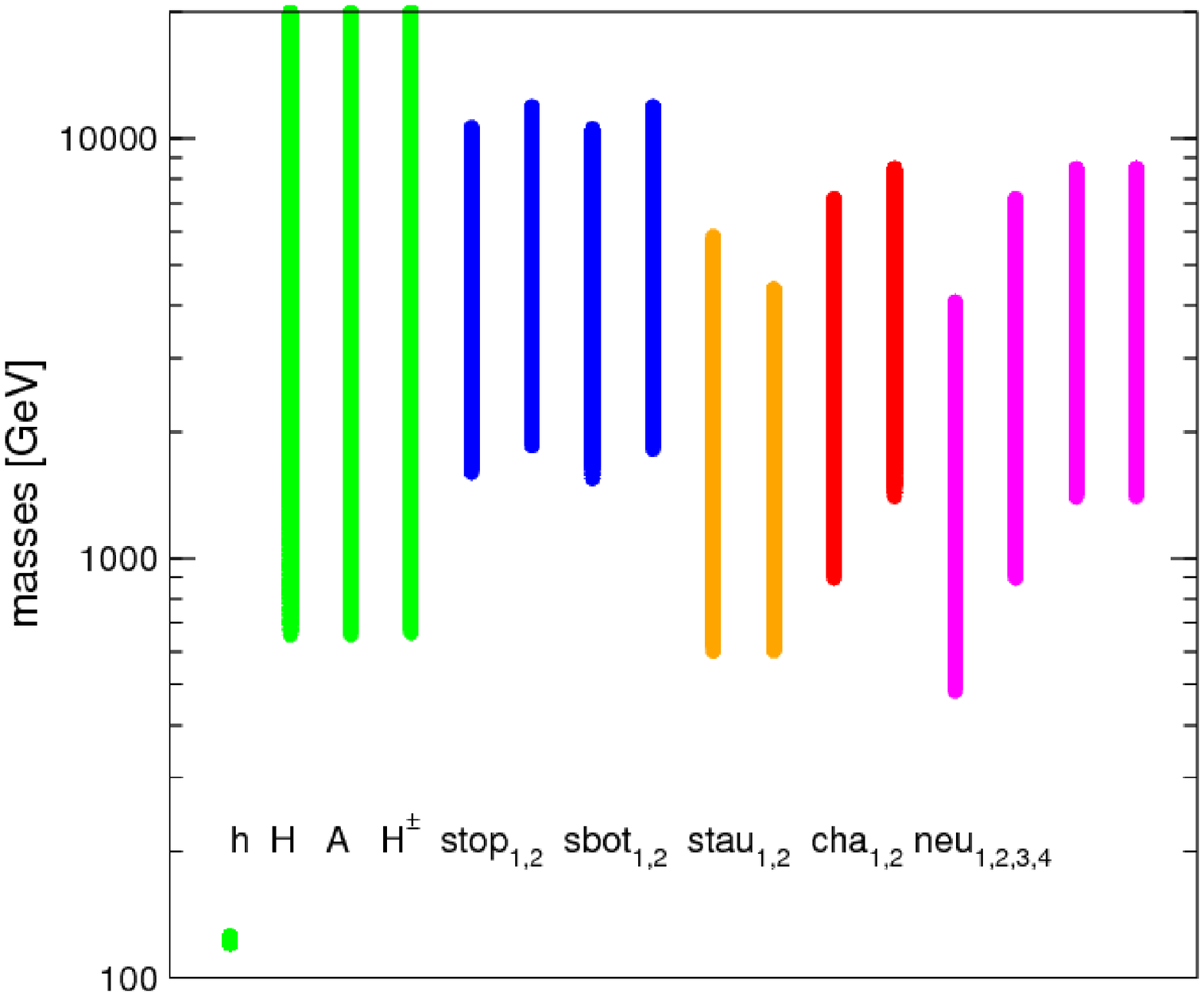}}
\vspace{2cm}
           \caption{The particle spectrum of model {\bf FUTB} with $\mu <0$,
             the points shown are in agreement with the quark mass
             constraints, the  
             $B$-physics observables, but the loose CDM constraint has been
             omitted (motivated by possible R-parity violation, see text).
  The color coding is as in \reffi{fig:masses}.}
\label{fig:masses2}
\vspace{-0.5em}
\end{figure}%
Finally for the model {\bf FUTB} with $\mu < 0$ we show the particle spectrum,
where only the quark mass constraints and the $B$ physics observables are
taken into account. The loose CDM constraint, on the other hand, has been
omitted, motivated by the possible R-parity violation as discussed above.
Consequently, following \reffi{fig:Higgs}, larger values of $M$ are allowed,
resulting in a heavier spectrum, as can be seen in \reffi{fig:masses2}. In
this case only part of the parameter space would be in the kinematic reach of
the LHC. Only the SM-like light Higgs boson remains observable for the whole
parameter range.


A more detailed analysis can be found in
\cite{Heinemeyer:2007tz,Heinemeyer:2010xt}.

\section{Finite ${SU(3)^3}$ model}
\subsection{Theoretical basis}

We now examine the possibility of constructing realistic FUTs based on
product gauge groups. Consider an $N=1$ supersymmetric theory, with
gauge group $SU(N)_1 \times SU(N)_2 \times \cdots \times SU(N)_k$,
with $n_f$ copies (number of families) 
of the
supersymmetric multiplets $(N,N^*,1,\dots,1) + (1,N,N^*,\dots,1) +
\cdots + (N^*,1,1,\dots,N)$.  The one-loop $\beta$-function
coefficient in the renormalization-group equation of each $SU(N)$
gauge coupling is simply given by
\begin{equation}
b = \left( -\frac{11}{3} + \frac{2}{3} \right) N + n_f \left( \frac{2}{3}
 + \frac{1}{3} \right) \left( \frac{1}{2} \right) 2 N = -3 N + n_f
N\,.
\label{3gen}
\end{equation}
This means that $n_f = 3$ is the only solution of \Eq {3gen} that
yields $b = 0$.  Since $b=0$ is a
necessary condition for a finite field theory, the existence of three
families of quarks and leptons is natural in such models, provided the
matter content is exactly as given above.

The model of this type with best phenomenology is the $SU(3)^3$ model
discussed in ref.~\cite{Ma:2004mi}, where the details of the model are
given. It corresponds to the well-known example of $SU(3)_C
\times SU(3)_L \times
SU(3)_R$~\cite{Derujula:1984gu,Lazarides:1993sn,Lazarides:1993uw,Ma:1986we},
 with quarks transforming as
\begin{equation}
  q = \begin{pmatrix} d & u & h \\ d & u & h \\ d & u & h \end{pmatrix}
\sim (3,3^*,1), ~~~ 
    q^c = \begin{pmatrix} d^c & d^c & d^c \\ u^c & u^c & u^c \\ h^c & h^c & h^c
\end{pmatrix} 
    \sim (3^*,1,3),
\label{2quarks}
\end{equation}
and leptons transforming as
\begin{equation}
\lambda = \begin{pmatrix} N & E^c & \nu \\ E & N^c & e \\ \nu^c & e^c & S 
\end{pmatrix}
\sim (1,3,3^*).
\label{3leptons}
\end{equation}
Switching the first and third rows of $q^c$ together with the first and 
third columns of $\lambda$, we obtain the alternative left-right model first 
proposed in ref.~\cite{Ma:1986we} in the context of superstring-inspired $E_6$. 

In order for all the gauge couplings to be equal at $M_{GUT}$, as is
suggested by the LEP results \cite{Amaldi:1991cn}, the cyclic symmetry $Z_3$ 
must be imposed, i.e.
\begin{equation}
q \to \lambda \to q^c \to q,
\label{15}
\end{equation}
where $q$ and $q^c$ are given in eq.~(\ref{2quarks}) and $\lambda$ in
eq.~(\ref{3leptons}).  Then, 
the first of the finiteness conditions (\ref{1st}) for one-loop
finiteness, namely the vanishing of the gauge $\beta$-function is
satisfied.

Next let us consider the second condition, i.e. the vanishing of the
anomalous dimensions of all superfields, eq.~(\ref{2nd}).  To do that
first we have to write down the superpotential.  If there is just one
family, then there are only two trilinear invariants, which can be
constructed respecting the symmetries of the theory, and therefore can
be used in the superpotential as follows
\begin{equation}
f ~Tr (\lambda q^c q) + \frac{1}{6} f' ~\epsilon_{ijk} \epsilon_{abc} 
(\lambda_{ia} \lambda_{jb} \lambda_{kc} + q^c_{ia} q^c_{jb} q^c_{kc} + 
q_{ia} q_{jb} q_{kc}),
\label{16}
\end{equation}
where $f$ and $f'$ are the Yukawa couplings associated to each invariant.
%
Quark and leptons obtain masses when the scalar parts of the
superfields $(\tilde N,\tilde N^c)$ obtain vacuum expectation values (vevs),
\begin{equation}
m_d = f \langle \tilde N \rangle, ~~ m_u = f \langle \tilde N^c \rangle, ~~ 
m_e = f' \langle \tilde N \rangle, ~~ m_\nu = f' \langle \tilde N^c \rangle.
\label{18}
\end{equation}

With three families, the most general superpotential contains 11 $f$
couplings, and 10 $f'$ couplings, subject to 9 conditions, due to the
vanishing of the anomalous dimensions of each superfield.  The
conditions are the following
\begin{equation}
\sum_{j,k} f_{ijk} (f_{ljk})^* + \frac{2}{3} \sum_{j,k} f'_{ijk}
(f'_{ljk})^* = \frac{16}{9} g^2 \delta_{il}\,,
\label{19}
\end{equation}
where
\begin{eqnarray}
&& f_{ijk} = f_{jki} = f_{kij}, \label{20}\\ 
&& f'_{ijk} = f'_{jki} = f'_{kij} = f'_{ikj} = f'_{kji} = f'_{jik}.
\label{21}
\end{eqnarray}
Quarks and leptons receive  masses when  the scalar part of the  
superfields $\tilde N_{1,2,3}$ and $\tilde N^c_{1,2,3}$ obtain vevs as follows
\begin{eqnarray}
&& ({\cal M}_d)_{ij} = \sum_k f_{kij} \langle \tilde N_k \rangle, ~~~ 
   ({\cal M}_u)_{ij} = \sum_k f_{kij} \langle \tilde N^c_k \rangle, \label{22} \\ 
&& ({\cal M}_e)_{ij} = \sum_k f'_{kij} \langle \tilde N_k \rangle, ~~~ 
   ({\cal M}_\nu)_{ij} = \sum_k f'_{kij} \langle \tilde N^c_k \rangle.
\label{23}
\end{eqnarray} 

We will assume that the below $M_{GUT}$ we have the usual MSSM, with
the two Higgs doublets coupled maximally to the third generation.
Therefore we have to choose
the linear combinations $\tilde N^c = \sum_i a_i \tilde N^c_i$ and
$\tilde N = \sum_i b_i \tilde N_i$ to play the role of the two Higgs
doublets, which will be responsible for the electroweak symmetry
breaking.  This can be done by choosing appropriately the masses in
the superpotential \cite{Leon:1985jm}, since they are not
constrained by the finiteness conditions.  We choose that
the two Higgs doublets are predominately coupled to the third
generation. Then these two Higgs doublets couple to the three
families differently, thus providing the freedom to understand 
their different masses and mixings.
The remnants of the $SU(3)^3$ FUT are the boundary conditions on the
gauge and Yukawa couplings, i.e. \eq{19}, the $h=-MC$
relation, and the soft scalar-mass sum rule eq.~(\ref{sumr}) at $M_{\rm
  GUT}$, which, when applied to the present model, takes the form
\begin{eqnarray}
m^2_{H_u} + m^2_{\tilde t^c} + m^2_{\tilde q} = M^2 = 
m^2_{H_d} + m^2_{\tilde b^c} + m^2_{\tilde q}~,
\end{eqnarray}
where   ${\tilde t^c}, ~{\tilde b^c}$, and ${\tilde q}$ are the
scalar parts of the corresponding 
superfields.

Concerning the solution to \eq{19} we consider two versions of the model:\\
I) An all-loop finite model with a unique and isolated solution, in
which $f'$ vanishes, which leads to the following relation
\begin{equation}
f^2 = f^2_{111} = f^2_{222} = f^2_{333} = \frac{16}{9} g^2\,.
\label{isosol}
\end{equation}
  As for
the lepton masses, because all $f'$ couplings have been fixed to be
zero at this order, in principle they would be expected to appear
radiatively induced by the scalar lepton masses appearing in the SSB
sector of the theory.  However, due to the finiteness
conditions they cannot appear radiatively and remain as a
problem for further study.  \\ 
II) A two-loop finite solution, in which we keep $f'$ non-vanishing
and we use it to introduce the lepton masses. The model in turn
becomes finite only up to two-loops since the corresponding solution
of \Eq{19} is not an isolated one any more, i.e. it is a parametric
one.  In this case we have the following boundary conditions for the
Yukawa couplings
\begin{eqnarray}
f^2 = r \left(\frac{16}{9}\right) g^2\,,\quad  
f'^2 = (1-r) \left(\frac{8}{3}\right) g^2\,,
\label{fprime}
\end{eqnarray}
where $r$ is a free parameter which parametrizes the different
solutions to the finiteness conditions. 
As for the boundary conditions of the soft scalars, we have the
universal case.

\subsection{Predictions for  ${SU(3)^3}$}\label{section6}

Below $M_{GUT}$ all couplings and masses of the theory run according
to the RGEs of the MSSM.  Thus we examine the evolution of these
parameters according to their RGEs up to two-loops for dimensionless
parameters and at one-loop for dimensionful ones imposing the
corresponding boundary conditions.  We further assume a unique
SUSY breaking scale $M_{SUSY}$ and below that scale the
effective theory is just the SM.

We compare our predictions with the experimental value $ m_t^{exp} =
(173.1 \pm 1.3)\,{\rm ~GeV}$ (again, if we use the most recent value
$m_t^{exp} = 173.3 \pm 1.1 \gev$~\cite{:1900yx} the results do not
change significantly), and recall that the theoretical values for
$m_t$ suffer from a correction of $\sim 4
\%$~\cite{Kubo:1997fi,Kobayashi:2001me,Mondragon:2003bp}. In the case
of the bottom quark, we take again the value evaluated at $M_Z$, $m_b
(M_Z)=2.83\pm 0.10\,{\rm GeV}$~\cite{Amsler:2008zzb}.  In the case of
model I, the predictions for the top quark mass (in this case $m_b$ is
an input) $m_t$ are $\sim 183\,{\rm ~GeV}$ for $\mu < 0 $, which is
above the experimental value, and there are no solutions for $\mu>0$.

\begin{figure}
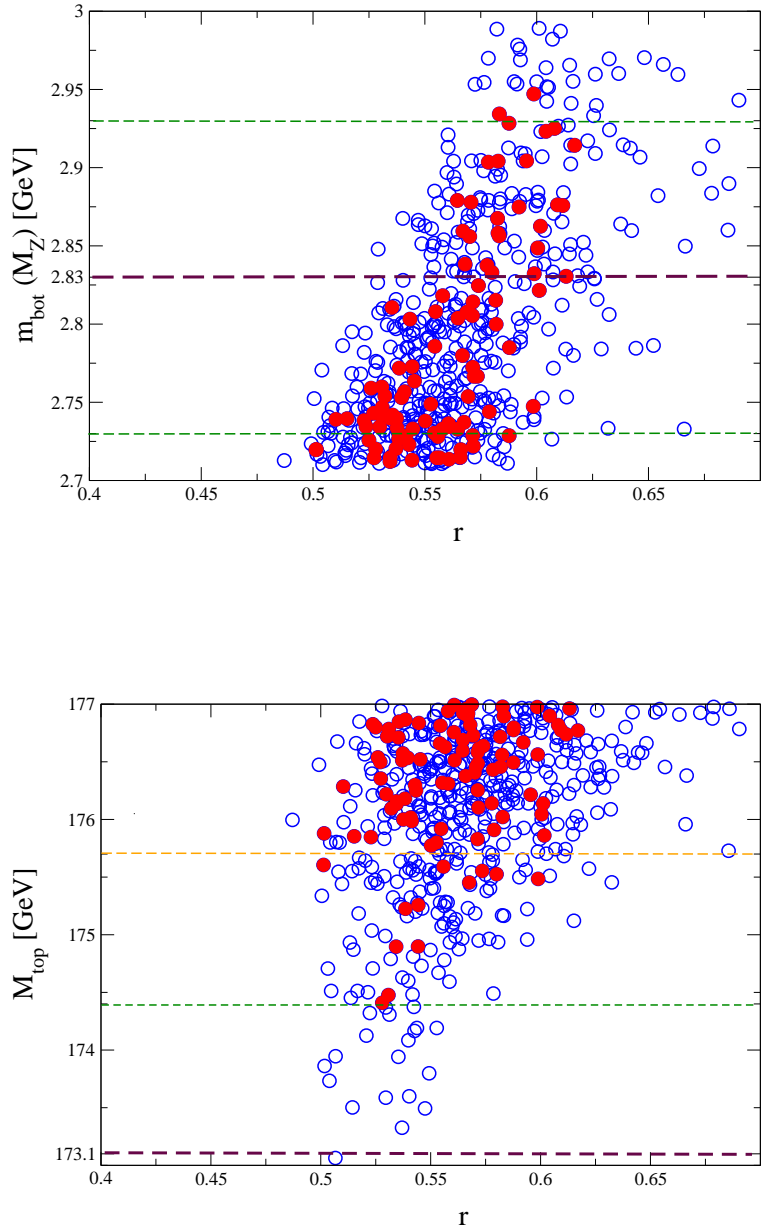

\begin{center}
\includegraphics[width=10cm,angle=0]{rvsMBOT-123.eps}\vspace{2cm}
\includegraphics[width=10cm,angle=0]{rvsMTOP-123.eps}
\caption{The figures show the values for the top and bottom quark
  masses for the FUT model $SU(3)^3$, with $\mu <0 $, vs the parameter
  $r$. The thicker horizontal line is the experimental central value,
  and the lighter green and orange ones are the one and two sigma
  limits respectively.  The red points are the ones that satisfy the B
  physics constraints.}
\label{fig:Higgs-m5}
\end{center}
\end{figure}

For the two-loop model {II}, we look for the values of the parameter
$r$ which comply with the experimental limits given above for top and
bottom quarks masses. In the case of $\mu >0$, for the bottom quark,
the values of $r$ lie in the range $0.15 \lesssim r \lesssim 0.32$.
For the top mass, the range of values for r is $0.35 \lesssim r
\lesssim 0.6$. From these values we can see that there is a very small
region where both top and bottom quark masses are in the experimental
range for the same value of $r$.  In the case of $\mu<0$ the situation
is similar, although slightly better, with the range of values $0.62
\lesssim r \lesssim 0.77$ for the bottom mass, and $0.4 \lesssim r
\lesssim 0.62$ for the top quark mass.  In the above mentioned analysis, the
masses of the new particles $h$'s and $E$'s of all families were taken
to be at the $M_{GUT}$ scale. Taking into account new thresholds for
these exotic particles below $M_{GUT}$ we  find a wider
phenomenologically viable parameter space.
This can be seen in Fig.\ref{fig:Higgs-m5}, where we took only one down-like
exotic particle decoupling at $10^{14}$ GeV, below that the usual MSSM. In
this case, for $r\sim 0.5 \sim 0.62$ we have reasonable agreement with
experimental data for both top and bottom quark masses, where the red points
in the figure are the ones that satisfy the B physics constraints \cite{Amsler:2008zzb}.

 The details of the
predictions of the $SU(3)^3$ are currently under a careful re-analysis in
view of the new value of the top-quark mass, the possible new
thresholds for the exotic particles, as well as different intermediate
gauge symmetry breaking into $SU(3)_c \times SU(2)_L \times SU(2)_R \times 
U(1)$ \cite{new-HMMZ}.

\section{Conclusions}
A number of proposals and ideas have matured with time and have
survived after careful theoretical studies and confrontation with
experimental data. These include part of the original GUTs ideas,
mainly the unification of gauge couplings and, separately, the
unification of the Yukawa couplings, a version of fixed point
behaviour of couplings, and certainly the necessity of SUSY as a way
to take care of the technical part of the hierarchy problem.  On the
other hand, a very serious theoretical problem, namely the presence of
divergencies in Quantum Field Theories (QFT), although challenged by
the founders of QFT \cite{Dirac:book,Dyson:1952tj,Weinberg:2009ca},
was mostly forgotten in the course of developments of the field partly
due to the spectacular successes of renormalizable field theories, in
particular of the SM. However, as it was already mentioned in the
Introduction, fundamental developments in Theoretical Particle Physics
are based in reconsiderations of the problem of divergencies and
serious attempts to solve it. These include the motivation and
construction of string and non-commutative theories, as well as $N=4$
supersymmetric field theories \cite{Mandelstam:1982cb,Brink:1982wv},
$N=8$ supergravity
\cite{Bern:2009kd,Kallosh:2009jb,Bern:2007hh,Bern:2006kd,Green:2006yu}
and the AdS/CFT correspondence \cite{Maldacena:1997re}.  It is a
thoroughly fascinating fact that many interesting ideas that have
survived various theoretical and phenomenological tests, as well as
the solution to the UV divergencies problem, find a common ground in
the framework of $N=1$ Finite Unified Theories, which we have
described in the previous sections. From the theoretical side they
solve the problem of UV divergencies in a minimal way. On the
phenomenological side, since they are based on the principle of
reduction of couplings (expressed via RGI relations among couplings
and masses), they provide strict selection rules in choosing realistic
models which lead to testable predictions. The celebrated success of
predicting the top-quark mass
\cite{Kapetanakis:1992vx,Mondragon:1993tw,Kubo:1994bj,Kubo:1994xa,Kubo:1995zg,Kubo:1996js}
is now extented to the predictions of the Higgs masses and the
supersymmetric spectrum of the MSSM
\cite{Heinemeyer:2007tz,Heinemeyer:2010xt}. At least the prediction
  of the lightest Higgs sector is expected to be tested in the next
  couple of years at LHC.

\subsection*{Acknowledgements}
It is a pleasure for one of us (G.Z.) to thank the Organizing
Committee for the very warm hospitality offered.  Similarly, many
thanks are due to the ITP Heidelberg, where this review was completed,
for the very warm hospitality. This work is partially supported by the
NTUA's programme supporting basic research PEBE 2009 and 2010, and the
European Union's ITN programme ``UNILHC'' PITN-GA-2009-237920. Supported
also by a mexican PAPIIT grant IN112709, and by Conacyt grants 82291
and 51554-F.

\small

\end{document}